\documentclass[runningheads]{llncs}
\usepackage[T1]{fontenc}
\usepackage{graphicx}
\usepackage{booktabs}
\usepackage[misc]{ifsym}
\newcommand{\corr}{(\Letter)}
\usepackage{mwe}
\usepackage{amsmath,amsfonts}
\usepackage{array}
\usepackage{subfig}
\usepackage{textcomp}
\usepackage{stfloats}
\usepackage{url}
\usepackage{verbatim}
\usepackage{graphicx}
\usepackage{balance}
\usepackage{hyperref}
\usepackage{eqnarray}
\usepackage{cite}
\usepackage{nicefrac}
\usepackage{bm}
\usepackage{amssymb}
\usepackage{mathtools}
\usepackage{bbm}
\usepackage{color}
\usepackage{caption}
\usepackage{algorithm}
\usepackage{appendix}
\usepackage[pdftex,dvipsnames]{xcolor}  % Coloured text etc.
\usepackage[colorinlistoftodos,prependcaption]{todonotes}
\usepackage{algpseudocode}
\usepackage{dsfont}
\usepackage{stackengine}

\newcommand{\biasedcomp}{{\textsf{BiasedComp}}}
\newcommand{\unbiasedcomp}{{\textsf{UnbiasedComp}}}
\newcommand{\compfedrl}{\textsf{CompFedRL}}

\newcommand{\topk}{\textsf{Top-$K$}}
\newcommand{\sparsifiedk}{\textsf{Sparsified-$K$}}
\newcommand{\compress}{\textsf{Compress}}

\usepackage[colorinlistoftodos]{todonotes}

\setstackEOL{\\}
\usepackage{etoolbox}
\usepackage{tikz}
\usetikzlibrary{tikzmark}
\usetikzlibrary{calc}
\errorcontextlines\maxdimen
\newcommand{\ALGtikzmarkcolor}{black}% customise this, if you want
\newcommand{\ALGtikzmarkextraindent}{4pt}% customise this, if you want
\newcommand{\ALGtikzmarkverticaloffsetstart}{-.5ex}% customise this, if you want
\newcommand{\ALGtikzmarkverticaloffsetend}{-.5ex}% customise this, if you want
\makeatletter
\newcounter{ALG@tikzmark@tempcnta}
\newcommand\ALG@tikzmark@start{%
    \global\let\ALG@tikzmark@last\ALG@tikzmark@starttext%
    \expandafter\edef\csname ALG@tikzmark@\theALG@nested\endcsname{\theALG@tikzmark@tempcnta}%
    \tikzmark{ALG@tikzmark@start@\csname ALG@tikzmark@\theALG@nested\endcsname}%
    \addtocounter{ALG@tikzmark@tempcnta}{1}%
}
\def\ALG@tikzmark@starttext{start}
\newcommand\ALG@tikzmark@end{%
    \ifx\ALG@tikzmark@last\ALG@tikzmark@starttext
    \else
        \tikzmark{ALG@tikzmark@end@\csname ALG@tikzmark@\theALG@nested\endcsname}%
        \tikz[overlay,remember picture] \draw[\ALGtikzmarkcolor] let \p{S}=($(pic cs:ALG@tikzmark@start@\csname ALG@tikzmark@\theALG@nested\endcsname)+(\ALGtikzmarkextraindent,\ALGtikzmarkverticaloffsetstart)$), \p{E}=($(pic cs:ALG@tikzmark@end@\csname ALG@tikzmark@\theALG@nested\endcsname)+(\ALGtikzmarkextraindent,\ALGtikzmarkverticaloffsetend)$) in (\x{S},\y{S})--(\x{S},\y{E});%
    \fi
    \gdef\ALG@tikzmark@last{end}%
}
\apptocmd{\ALG@beginblock}{\ALG@tikzmark@start}{}{\errmessage{failed to patch}}
\pretocmd{\ALG@endblock}{\ALG@tikzmark@end}{}{\errmessage{failed to patch}}
\makeatother
\begin{document}
\title{Compressed Federated Reinforcement Learning with a Generative Model}
\titlerunning{CompFedRL with a Generative Model}
% If the full title of your paper is short enough to also fit in the running head, you can omit the abbreviated paper title here. You can check as follows: if you comment out the \titlerunning line, something will appear in the header of all odd-numbered pages of your PDF from page 3 onward. This something is either the full title (in which case all is well), or the error message "Title Suppressed Due to Excessive Length". If this error message appears, you're going to want to provide an abbreviated title within the \titlerunning command, because if you won't do it, Springer will do it for you.

%N.B.: Author information (both in the \author{} and \authorrunning{} command) should only be present in the Camera-Ready Version of your paper. The version that you initially submit for review, ought to be double-blind. So, when initially submitting your paper, use:
%\author{Author information scrubbed for double-blind reviewing}
\author{Ali Beikmohammadi\inst{1} \corr \orcidID{0000-0003-4884-4600} \and Sarit Khirirat\inst{2}\orcidID{0000-0003-4473-2011}\footnote{Sarit Khirirat contributed to this paper before he joined King Abdullah University of Science and Technology (KAUST), Thuwal, Saudi Arabia. He is currently a postdoctoral fellow at KAUST.} \and Sindri Magn\'usson\inst{1}\orcidID{0000-0002-6617-8683}}

\authorrunning{A. Beikmohammadi et al.}

\institute{Department of Computer and Systems Sciences, \\
  Stockholm University, SE-164 25 Stockholm, Sweden \\
\email{\{beikmohammadi, sindri.magnusson\}@dsv.su.se}
\and
King Abdullah University of Science and Technology (KAUST), \\
Thuwal, Saudi Arabia \\
\email{sarit.khirirat@kaust.edu.sa}
}

\toctitle{Compressed Federated Reinforcement Learning with a Generative Model} 
\tocauthor{Ali~Beikmohammadi,~Sarit~Khirirat,~Sindri~Magn\'usson}

\maketitle              % typeset the header of the contribution

\begin{abstract}
Reinforcement learning has recently gained unprecedented popularity, yet it still grapples with sample inefficiency. Addressing this challenge, federated reinforcement learning (FedRL) has emerged, wherein agents collaboratively learn a single policy by aggregating local estimations. However, this aggregation step incurs significant communication costs.
In this paper, we propose \compfedrl, a communication-efficient FedRL approach incorporating both \emph{periodic aggregation} and \emph{(direct/error-feedback) compression} mechanisms. Specifically, we consider compressed federated $Q$-learning with a generative model setup, where a central server learns an optimal $Q$-function by periodically aggregating compressed $Q$-estimates from local agents. For the first time, we characterize the impact of these two mechanisms (which have remained elusive) by providing a finite-time analysis of our algorithm, demonstrating strong convergence behaviors when utilizing either direct or error-feedback compression. Our bounds indicate improved solution accuracy concerning the number of agents and other federated hyperparameters while simultaneously reducing communication costs. To corroborate our theory, we also conduct in-depth numerical experiments to verify our findings, considering \textsf{Top-$K$} and \textsf{Sparsified-$K$} sparsification operators.

\keywords{Federated Reinforcement Learning  \and Communication Efficiency \and Direct Compression \and Error-feedback Compression.} % \and Top-$K$ Sparsification \and Sparsified-$K$ Sparsification
\end{abstract}
\section{Introduction}

Reinforcement learning (RL) \cite{sutton2018reinforcement} has advanced many progresses in solving several decision-making applications, such as game playing \cite{mnih2015},
%financial markets \cite{fischer2018},
%robotic control \cite{haarnoja2018soft},
optimal control \cite{beikmohammadi2023ta},
%healthcare \cite{CORONATO2020101964},
autonomous driving \cite{ref4},
and recommendation systems~\cite{rec1}. 
In traditional RL algorithms, an agent learns and makes the best sequential actions within a Markov Decision Process (MDP) that maximize long-term rewards, even without the full knowledge of its model parameters. 
This learning process involves the agent repeatedly interacting with the environment, generating data that can then be used to refine its decision-making strategy.

However, training RL models under high-dimensional environments is challenging, because the RL models require (1) high volumes of storage being directly proportional to the size of state and action spaces, and (2) high sample complexity to find an optimal policy, i.e. a large number of interactions between the agent and the environment \cite{beikmohammadi2024accelerating, beikmohammadi2023human}.
%\cite{beikmohammadi2024accelerating, beikmohammadi2023human}.
Training RL models with limited data often leads to poor performance in high output variability, while sequentially generating vast volumes of data is computationally expensive. 
Consequently, many practical RL implementations resort to parallel data sampling from the environment by multiple agents \cite{mnih2016asynchronous}. 

Recently, many RL algorithms have been modified to leverage federated learning (FL), known as federated RL (FedRL), to obtain faster training performance and improved accuracy. 
FL is a popular multi-agent framework that ensures privacy and security~\cite{mcmahan2017communication} since it does not require the agents to transmit their private data over the network, which is computationally prohibitive, especially when the data is sizeable, and the network is resource-constrained. 
At each communication round of FedRL algorithms, the central server updates the model parameters, based on aggregations of local model updates transmitted by all the agents. 
Previous studies on FedRL algorithms primarily focus on addressing challenges such as robustness to adversarial attacks \cite{wu2021byzantine, fan2021fault}, 
environment heterogeneity \cite{jin2022federated, wang2023federated, zhang2024finite}, 
behavior policy heterogeneity \cite{woo2023blessing}, 
and sample complexities \cite{doan2021finite, khodadadian2022federated}.
Unfortunately, none of these studies addresses the communication performance bottleneck of FedRL algorithms that interact with high-dimensional environments. %This highlights an important gap in current research efforts. 

A commonly used approach to reduce communication costs is to apply compression, e.g., sparsification and quantization, alongside periodic aggregation.
Compressing the information reduces its size before it is transmitted and is beneficial for running optimization algorithms over bandwidth-limited networks. %to train huge-dimensional state-of-the-art learning models (e.g. ResNet \cite{he2016deep,alistarh2017qsgd}).  
The impact of compression on optimization algorithms under both centralized and federated settings has been extensively studied, e.g. in \cite{alistarh2017qsgd,alistarh2018convergence,wang2018atomo,stich2018sparsified,khirirat2020compressed,beznosikov2023biased}. 
Unfortunately, very limited works have studied compressed FedRL algorithms. 
Multi-agent RL with compressed updates 
%under both i.i.d. and Markovian sampling 
has been studied in \cite{mitra2023temporal}. 
However, the result in~\cite{mitra2023temporal} is limited to TD learning with linear function approximation, and they neither utilized periodic averaging nor analyzed direct compression.  

%Although the convergence of multi-agent TD(0) learning with compressed updates under both i.i.d. and Markovian sampling has been studied by \cite{mitra2023temporal}, the results are limited to TD(0) learning and 
%are still limited to the settings where multiple agents interact with the same MDP and collaborate to evaluate the same policy, but access to different data realizations. 

%\subsection{Contributions}
\paragraph{Contributions.}
\begin{itemize}\setlength\itemsep{0.5em}
    %\item \textbf{Communication-Efficient FedRL.} In Section \ref{compfedrl}, we propose \compfedrl, a novel approach to FedRL designed to mitigate the substantial communication costs associated with the aggregation step. Our method incorporates both \emph{periodic aggregation} and \emph{(direct/error-feedback) compression} mechanisms.
    \item \textbf{\compfedrl}. In Section \ref{compfedrl}, we propose \compfedrl, a novel FedRL algorithm that incorporates both \emph{periodic aggregation} and \emph{(direct/error-feedback) compression} mechanisms to guarantee strong convergence while mitigating substantial costs of communicating local $Q$-estimates. 
    \item \textbf{Non-asymptotic convergence results for \compfedrl ~under \unbiasedcomp ~and \biasedcomp ~compression.} We provide the convergence analysis of \compfedrl, characterizing the impact of these mechanisms on convergence behaviors.
    Through a finite-time analysis of our algorithm, we demonstrate robust convergence behaviors when employing either direct or error-feedback compression techniques in Section~\ref{sec:convergence}. Our theoretical bounds highlight the enhanced solution accuracy achieved by \compfedrl, particularly concerning the number of agents and other federated hyperparameters, while simultaneously reducing communication costs.
    \item \textbf{Empirical validation.} To validate our theoretical observations, in Section \ref{Experiments}, we conduct extensive numerical experiments, leveraging sparsification operators such as \topk ~and \sparsifiedk. These experiments corroborate the effectiveness of \compfedrl ~in practice, further reinforcing the significance of our contributions in advancing communication-efficient FedRL algorithms.
    Our experiments’ code is publicly available in the following link: \url{https://github.com/AliBeikmohammadi/CompFedRL}.
\end{itemize}

\section{Related Work}
In what follows, we delve into the most pertinent threads of related work, with a particular focus on theoretical contributions in the field.
We present prior works closely related to our paper in RL and communication-efficient learning.

\subsection{Single-agent RL Algorithms}
%\paragraph{Single-agent RL Algorithms.}
%\paragraph{Analysis of Single-agent RL Algorithms.}
Existing works have extensively investigated the convergence guarantees of many RL algorithms, such as TD-learning and $Q$-learning algorithms, primarily on the single-agent environment under various conditions. 
TD-learning algorithms are shown to have asymptotic convergence in the on-policy setting by \cite{tsitsiklis1996analysis, tadic2001convergence, borkar2009stochastic} and in the off-policy setting by \cite{maei2018convergent, zhang2020provably}, while the algorithms attain non-asymptotic convergence (or finite-sample bounds) in the on-policy setting by \cite{dalal2018finite, lakshminarayanan2018linear, bhandari2018finite, srikant2019finite, hu2019characterizing, chen2021lyapunov} and in the off-policy setting by \cite{chen2020finite, chen2021lyapunov}.
Furthermore, initially proposed by \cite{watkins1992q},
the $Q$-learning algorithm %has also been subject to significant research efforts aimed at establishing its convergence properties. 
has been extensively studied its convergence properties in varied settings. 
For instance, on the one hand, the asymptotic convergence of the $Q$-learning algorithm was established by \cite{tsitsiklis1994asynchronous, jaakkola1993convergence, bertsekas1996neuro, szepesvari1997asymptotic, borkar2000ode, borkar2009stochastic}.
On the other hand, the finite-time convergence of $Q$-learning algorithm under synchronous sampling, also known as RL with a generative model, was proved by  \cite{even2003learning, beck2012error, wainwright2019stochastic, chen2020finite, li2024q, gheshlaghi2013minimax}, and under asynchronous sampling (known as Markovian sampling) was studied by \cite{even2003learning, beck2012error, qu2020finite, li2024q, li2020sample, chen2021lyapunov}.
Our work operates within RL with a generative model, but focuses on a multi-agent collaborative federated setup, emphasizing communication-efficient methodologies, which requires an analysis framework different from traditional single-agent RL analyses.

%In this paper, we investigate communication-efficient FedRL with a generative model. We derive non-asymptotic convergence results, which require an analysis framework different from traditional single-agent RL algorithms. 

%Our work also falls within the realm of RL with a generative model setup. However, we consider a multi-agent collaborative federated setup, focusing on communication-efficient methodologies. This adds complexity to our study, particularly in investigating non-asymptotic results, which differ significantly from traditional single-agent RL analyses.

\subsection{Distributed and Federated RL Algorithms}
%\paragraph{Distributed and Federated RL Algorithms.}
%\paragraph{Analysis of Distributed and Federated RL Algorithms.}
Several recent studies have introduced distributed variants of RL algorithms aimed at expediting training processes \cite{mnih2016asynchronous, espeholt2018impala,assran2019gossip}. 
Theoretical examinations of the convergence properties of these distributed RL algorithms have also been explored in recent literature across various scenarios, including decentralized stochastic approximation \cite{doan2019finite, sun2020finite, wai2020convergence, zheng2023federated, mitra2023temporal}, TD learning with linear function approximation \cite{wang2023federated}, and off-policy TD in actor-critic algorithms \cite{chen2021multi}. 
Additionally, \cite{chen2022sample, shen2023towards} have conducted analyses on the finite-time convergence of distributed actor-critic algorithms. 
Nevertheless, all of these investigations focus on decentralized setups, where agents communicate with their neighbors following each local update.
On the other hand, there are some studies that consider a federated setting, with each agent performing multiple local updates between successive communication rounds, thereby resulting in communication savings. This includes TD learning with linear function approximation \cite{dal2023federated, dal2023over, khodadadian2022federated}, off-policy TD learning \cite{khodadadian2022federated}, SARSA with linear function approximation \cite{zhang2024finite}, and $Q$-learning \cite{khodadadian2022federated, woo2023blessing, jin2022federated}.
Our work, akin to the setup of \cite{woo2023blessing}, centers on a federated $Q$-learning algorithm but sets itself apart by integrating various compression strategies to enhance communication efficiency.

%Our work aligns with the setup of \cite{woo2023blessing}, focusing on a federated $Q$-learning algorithm. However, we distinguish ourselves by incorporating different compression strategies, including direct and error-feedback compression, to achieve communication efficiency.

\subsection{Communication-Efficient Learning Algorithms}
%\paragraph{Communication-Efficient Learning Algorithms.}
%\paragraph{Communication-Efficient Algorithms for Distributed and Federated Learning.}
Many compression schemes, i.e. sparsification and/or quantization, have been used for optimization algorithms to train state-of-the-art learning models (e.g. AlexNet, VGG19, and ResNet152) under resource-constrained networks \cite{alistarh2017qsgd, wen2017terngrad, wang2018atomo, mayekar2020ratq}. 
These compression schemes can be either \emph{unbiased} (such as 
QSGD \cite{alistarh2017qsgd}, TernGrad \cite{wen2017terngrad}, and ATOMO \cite{wang2018atomo})
or \emph{biased} (e.g. \topk \cite{alistarh2018convergence, beikmohammadi2024distributed}, normalized gradient \cite{mandic2004generalized} and sign compression \cite{bernstein2018signsgd}). 
Despite communication benefits, optimization algorithms with compression often have poor convergence performance.
For instance, the biased compression algorithms are shown to fail to converge for simple problems \cite{beznosikov2023biased,khirirat2020compressed}. 
%
%
%Over the past decade, various sparsification schemes have been investigated for optimization tasks , gaining popularity in the realm of deep learning.
%
%
%Alongside the success of unbiased sparsification operators in compression techniques, known as direct or unbiased compression, methods like \topk, which involve transmitting only a subset of gradient vector components with the largest magnitudes, have demonstrated empirical benefits of extreme sparsification \cite{aji2017sparse, lin2017deep}.
%However, establishing theoretical guarantees for such techniques presents challenges due to their inherently biased nature.
To address the problem of direct compression, several studies \cite{stich2018sparsified, khirirat2020compressed, alistarh2018convergence, karimireddy2019error, lin2022differentially, gorbunov2020linearly, mitra2021linear, beikmohammadi2024convergence} have proposed error-feedback mechanisms which leverage memory of previous compression errors to mitigate compression biases for distributed and federated learning algorithms under various settings. %Recently, Beznosikov et al. \cite{beznosikov2023biased} and Gorbunov et al. \cite{} have provided theoretical insights into biased sparsification within a distributed architecture, while Mitra et al. \cite{mitra2021linear} explored sparsification in the context of federated learning.
None of the mentioned works has investigated communication-efficient learning in the context of RL. The only existing study, \cite{mitra2023temporal}, examines error-feedback compression, focusing on TD learning with linear function approximation in distributed learning, contrasting with our synchronous federated $Q$-learning approach, where aggregation occurs periodically. We investigate both direct and error-feedback compression operators, offering a thorough FedRL convergence analysis.

%However, there is a very limited number of works that investigate RL algorithms for communication-efficient learning. For instance, only one work by \cite{mitra2023temporal} incorporated error-feedback schemes into multi-agent TD(0) learning with linear function approximation. Their convergence results are limited to the setting where all agents interact with the same MDP and evaluate the same policy, but different state transitions. In this work, we consider a synchronous $Q$-learning for a federated learning setup, which is more general than \cite{mitra2023temporal}. In particular, we explore the impact of direct and error-feedback compression on the convergence performance of algorithms. Our results show the speed-up with respect to the number of agents, like \cite{mitra2023temporal}. 

%Yet, none of the aforementioned works has explored the impact of biased or unbiased compression in the realm of RL. The only existing work in RL that has recently considered error-feedback compression is \cite{mitra2023temporal}. Nevertheless, their study focuses on TD learning with linear function approximation in distributed learning, which differs substantially from our work. Specifically, we consider a synchronous $Q$-learning in a federated setup, where aggregation occurs periodically. Moreover, we explore the effectiveness of both direct and error-feedback compression operators, providing a comprehensive analysis of communication-efficient algorithms in the context of FedRL.

\subsection{Notation}
%\paragraph{Notation.}
%\ali{Please add (remove) any notation you feel should (not) be clarified.}
%\sarit{Should be complete now. Since I use $\ell_\infty$-norm and $[a,b]$.}
We denote by $\mathbb{N}$, $\mathbb{N}_0$, and $\mathbb{R}$ the sets of natural numbers, natural numbers including zero, and real numbers, respectively. The set $\{a,a+1, ..., b\}$ is represented as $[a, b]$, where $a, b \in \mathbb{N}_0$ and $a \leq b$. 
For a vector $v \in \mathbb{R}^d$, 
%$\| v \|_1$, 
$\| v \|_2$ and $\| v\|_\infty$ are its 
%$\ell_1$-norm, 
$\ell_2$-norm and $\ell_\infty$-norm, respectively.

\section{Preliminaries and Background}
Within this section, we present the mathematical framework and fundamental concepts of MDPs.

\subsection{Discounted Infinite-horizon MDP}
%\paragraph{Discounted Infinite-horizon MDP.}
We consider a discounted infinite-horizon MDP, denoted by a quintuple $\mathcal{M}=(\mathcal{S}, \mathcal{A}, P, r, \gamma)$, where $\mathcal{S} = \{1,2,..., \vert \mathcal{S}\vert\}$ and $\mathcal{A}= \{1,2,..., \vert \mathcal{A}\vert\}$ represent the finite state space and action space, respectively. The transition probability kernel $P: \mathcal{S} \times \mathcal{A} \times \mathcal{S} \rightarrow[0,1]$ determines the likelihood $P\left(s^{\prime} \mid s, a\right)$ of transitioning from state $s$ to state $s^{\prime}$ upon taking action $a$. The reward function $r: \mathcal{S} \times \mathcal{A} \rightarrow [-1,1]$ assigns an immediate reward $r(s, a)$ for executing action $a$ in state $s$. Additionally, the discount factor $\gamma \in (0,1)$ adjusts the weighting of future rewards.

\subsection{Policy, and $Q$-function}
%\paragraph{Policy, and $Q$-function.}
A policy $\pi$ is a mapping $\pi: \mathcal{S} \rightarrow \Delta(\mathcal{A})$, where $\pi(a \mid s)$ denotes the probability of selecting action $a$ given state $s$. Given a policy $\pi$, the value function $V^{\pi}: \mathcal{S} \rightarrow \mathbb{R}$ quantifies the expected discounted cumulative reward starting from an initial state $s$ and is defined as
\begin{equation}
V^{\pi}(s)=\mathbb{E}\left[\sum_{k=0}^{\infty} \gamma^{k} r\left(s_{k}, a_{k}\right) \mid s_{0}=s\right], \forall ~ s \in \mathcal{S}. 
\end{equation}
Here, the expectation is computed over the randomness of the trajectory $\left\{s_{k}, a_{k}, r_{k}\right\}_{k=0}^{\infty}$, sampled according to the transition kernel (i.e., $s_{k+1} \sim P\left(\cdot \mid s_{k}, a_{k}\right)$ ) and policy $\pi$ (i.e., $a_{k} \sim \pi\left(\cdot \mid s_{k}\right)$ ) for any $k \geq 0$. Similarly, the state-action value function, denoted as the $Q$-function $Q^{\pi}: \mathcal{S} \times \mathcal{A} \rightarrow \mathbb{R}$, measures the expected discounted cumulative reward from an initial state-action pair $(s, a)$:
\begin{equation}
Q^{\pi}(s, a)=\mathbb{E}\left[\sum_{k=0}^{\infty} \gamma^{k} r\left(s_{k}, a_{k}\right) \mid s_{0}=s, a_{0}=a\right], \forall ~(s, a) \in \mathcal{S} \times \mathcal{A}.
\end{equation}
\subsection{Optimal Policy and  Bellman Operator}
%\paragraph{Optimal Policy and  Bellman Operator.}
An optimal policy $\pi^{\star}$ maximizes the value function uniformly across all states. Note that the existence of such an optimal policy is always assured \cite{puterman2014markov, sutton2018reinforcement}, and it simultaneously maximizes the $Q$-function. The corresponding optimal value function and $Q$-function are represented as $V^{\star}=V^{\pi^{\star}}$ and $Q^{\star}=Q^{\pi^{\star}}$, respectively. 
It is a well-established fact that the optimal $Q$-function $Q^{\star}$ can be determined as the unique fixed point of the population Bellman operator $\mathcal{T}: \mathbb{R}^{\vert \mathcal{S}\vert \times \vert \mathcal{A}\vert} \rightarrow \mathbb{R}^{\vert \mathcal{S}\vert \times \vert \mathcal{A}\vert}$, which is defined as:
\begin{equation} \label{eq:exactbellman}
\mathcal{T}(Q)(s, a)=r(s, a)+\gamma \underset{s^{\prime} \sim P(\cdot \mid s, a)}{\mathbb{E}}\left[\max _{a^{\prime} \in \mathcal{A}} Q\left(s^{\prime}, a^{\prime}\right)\right].
\end{equation}

\subsection{RL with a Generative Model}
%\paragraph{RL with a Generative Model.}
In the context of learning, the transition kernel $P$ is typically unknown. Thus computing the Bellman operator precisely is infeasible. 
%making it infeasible to evaluate the Bellman operator precisely. 
Instead, we assume access to a simulation engine that generates samples. In this paper, we focus on the \emph{synchronous} or \emph{generative} setting, where at each time step $k$ and for each state-action pair $(s,a)$, a sample $s_k(s,a)$ is observed, drawn according to the transition function $P(\cdot \mid s, a)$ (i.e., $s_k(s,a) \sim P(\cdot \mid s, a)$ for all $(s, a) \in \mathcal{S} \times \mathcal{A}$). 
%
%It's worth noting that guarantees 
Also, convergence guarantees 
for the synchronous setting can be easily translated %into guarantees 
for the \emph{asynchronous} setting, e.g. by using cover times of Markov chains \cite{even2003learning, azar2011speedy}. 
%conversions of this type.

In the synchronous form of the $Q$-learning algorithm, a sequence of $Q$-tables $\left\{Q_{k}\right\}_{k \geq 1}$ is generated according to:
\begin{equation} \label{eq:SyncQnormal}
Q_{k+1}=\left(1-\eta\right) Q_{k}+\eta \mathcal{T}_{k}\left(Q_{k}\right),
\end{equation}
where $\eta \in (0,1]$ is a learning rate, and the operator $\mathcal{T}_{k}: \mathbb{R}^{|\mathcal{S}| \times|\mathcal{A}|} \rightarrow \mathbb{R}^{|\mathcal{S}| \times|\mathcal{A}|}$ is an empirical Bellman operator with its $(s, a)$-entry evaluated by:
\begin{equation} \label{eq:empricalbellman}
\mathcal{T}_{k}(Q)(s, a)=r(s, a)+\gamma \max _{a^{\prime} \in \mathcal{A}} Q\left(s_{k}, a^{\prime}\right),
\end{equation}
where $s_{k} \in \mathbb{R}^{|\mathcal{S}| \times|\mathcal{A}|}$ is a random matrix indexed by state-action pairs $(s, a)$, with each entry $s_{k}(s, a)$ drawn from the probability distribution $P(\cdot \mid s, a)$. 
Note that $\mathbb{E}\left[\mathcal{T}_{k}(Q)\right]=\mathcal{T}(Q)$ for any fixed $Q$, indicating that the empirical Bellman operator \eqref{eq:empricalbellman} is an unbiased estimate of the population Bellman operator \eqref{eq:exactbellman}.

\section{\compfedrl
%Communication-efficient FedRL with a Generative Model
} \label{compfedrl}
\begin{algorithm}[t] %htp!
	%\caption{Compressed-efficient Synchronous $Q$-learning (\compfedrl)}\label{alg:EFFedQSync}
        \caption{Compressed FedRL with a Generative Model (\compfedrl)} \label{alg:EFFedQSync}
	\begin{algorithmic}[1]
		\State \textbf{inputs:} Learning rate $\eta \in (0,1]$, Discount factor $\gamma \in (0,1)$, Federated parameter $\beta \in (0,1]$, 
		 Number of agents $I$, Number of local epochs $K$, Number of communication rounds $T$, \compress ~operator $\in \{\unbiasedcomp,  \biasedcomp\}$.
		\State \textbf{initialization:} Server broadcasts $\overline Q_{0}=Q_{0}$.
            \If{\compress ~operator is \biasedcomp}
            \State Initial errors $e_{0}^{(i)}=0, \forall i \in [I]$.
            \EndIf
		\For{each round $t=0,\ldots,T-1$}
		\For{each agent $i \in [I]$ \textbf{in parallel}}
		\State Agent $i$ initializes $Q_{t,0}^{(i)} = \overline Q_{t}$ 
		\For{$k=0,\ldots,K-1$}
		\State Draw $s_{t,k+1}^{(i)} \left(s,a\right)  \sim {P}{\left(.|s,a\right)}$ for all ${\left(s,a\right)} \in \mathcal{S} \times \mathcal{A}$.
		\State Update local $Q_{t,k+1}^{(i)}$ according to \eqref{standardFedQSync}
		\EndFor 
            \If{\compress ~operator is \unbiasedcomp}
            \State Compute $h_{t}^{(i)} = \compress(Q_{t,K}^{(i)}-\overline Q_{t} )$ and send back to the server.
            \EndIf
            \If{\compress ~operator is \biasedcomp}
            \State Compute $h_{t}^{(i)} = \compress(Q_{t,K}^{(i)}-\overline Q_{t} + e_{t}^{(i)})$ and send back to the server.
            \State Update $e_{t+1}^{(i)} = Q_{t,K}^{(i)}-\overline Q_{t} + e_{t}^{(i)} -  h_{t}^{(i)}$.
            \EndIf
		\EndFor 
		\State Server computes and broadcasts global $\overline{Q}_{t+1} = \overline Q_{t} + \frac{ \beta}{I}\sum_{i=1}^{I} {h_{t}^{(i)}}$
		\EndFor
	\end{algorithmic}
\end{algorithm}
We introduce communication-efficient FedRL with a generative model called \compfedrl. 
This algorithm utilizes periodic aggregation and compression mechanisms to obtain provably strong convergence guarantees while reducing communication bandwidth for transmitting local $Q$-estimates from the agents to the central server. 
At each communication round $t=[0,T-1]$ of the algorithm, every agent $i\in[1,I]$ receives a global $Q$-table $\overline Q_t$ from the central server and initializes its $Q$-estimate table $Q^{(i)}_{t,0}=\overline Q_t$. 
Each agent in parallel updates all entries of its $Q$-estimates for $K$ local epochs according to: 
\begin{equation} \label{standardFedQSync}
	Q_{t,k+1}^{(i)}\left(s,a\right) = \left(1-\eta \right)Q_{t,k}^{(i)}\left(s,a\right)+\eta  \mathcal{T}^{(i)}_{k}(Q_{t,k}^{(i)}) \left(s,a\right) 
\end{equation}
where $\eta$ is a positive learning rate, and $\mathcal{T}^{(i)}_{k}$ is the empirical Bellman operator at local epoch $k$ defined by
\begin{equation} \label{eq:DefSyncBellman}
	\mathcal{T}^{(i)}_{k}(Q)\left(s,a\right)=r\left(s,a\right)+\gamma \max _{a^{\prime} \in \mathcal{A}} Q\left(s_{t,k+1}^{(i)} \left(s,a\right), a^{\prime}\right), \quad s_{t,k+1}^{(i)} \left(s,a\right)\sim P(\cdot | s,a). 
\end{equation}

After running for $K$ local epochs, each agent transmits to the server the compressed progress between the local $Q$-estimate and the global $Q$-table denoted by $h_t^{(i)}$. Here, $h_t^{(i)} = \compress(Q_{t,K}^{(i)}-\overline Q_{t} )$ if $\compress(\cdot)$ is \unbiasedcomp, and $h_t^{(i)} = \compress(Q_{t,K}^{(i)}-\overline Q_{t} + e_t^{(i)}) $ if $\compress(\cdot)$ is \biasedcomp~where
%\begin{eqnarray*}
%	h_t^{(i)} = \begin{cases}  \compress(Q_{t,K}^{(i)}-\overline Q_{t} ) & \text{for \unbiasedcomp} \\
%		\compress(Q_{t,K}^{(i)}-\overline Q_{t} + e_t^{(i)}) & \text{for \biasedcomp} 	 \end{cases}
%\end{eqnarray*}	
$e_t^{(i)}$ maintains the memory of compression errors defined by $e_{t+1}^{(i)} = Q_{t,K}^{(i)}-\overline Q_{t} + e_{t}^{(i)} -  h_{t}^{(i)}$. 
%\begin{eqnarray*}
%	e_{t+1}^{(i)} = Q_{t,K}^{(i)}-\overline Q_{t} + e_{t}^{(i)} -  h_{t}^{(i)}.
%\end{eqnarray*}	

Finally, after receiving all $h_t^{(i)}$ from every agent, the central server updates the global $Q$-table $\overline Q_{t+1}= \overline Q_t + ({\beta}/{I}) \sum_{i=1}^I h_t^{(i)}$, 
%defined by:
%\begin{eqnarray*}
%	\overline Q_{t+1} = \overline Q_t + \frac{\beta}{I} \sum_{i=1}^I h_t^{(i)},
%\end{eqnarray*}	
where $\beta \in [0,1]$ is the federated parameter.
The full description of \compfedrl~ is in Algorithm~\ref{alg:EFFedQSync}.
Further note that Algorithm~\ref{alg:EFFedQSync} becomes QAvg \cite{jin2022federated} and FedSyncQ \cite{woo2023blessing} when  $\compress(v) = v$ and $\beta=1$, and recovers a centralized synchronous $Q$-learning algorithm analyzed by~\cite{li2024q,beck2012error,chen2020finite,even2003learning} by choosing $\compress(v) = v$, $\beta=1$ and $I=1$.

\subsection{Compression Options for \compfedrl}

The choices of compression operators deployed in \compfedrl~are flexible. 
We define two types of compression operators that are useful for our analysis.%: \unbiasedcomp~and \biasedcomp.
\begin{definition}[\unbiasedcomp]\label{def:unbiasedcomp}
$\compress(v):\mathbb{R}^d\rightarrow\mathbb{R}^d$ is \unbiasedcomp~if it satisfies the following: for $j\in [1,d]$ and $v\in\mathbb{R}^d$ (1) $\mathbf{E} [\compress(v)] = v$, (2) $\mathbf{E} \vert   \compress(v)_j - v_j\vert^2 \leq q_2 \vert v_j \vert^2$ for some $q_2 >0$, and (3) $\| \compress(v) - v \|_\infty \leq q_\infty \| v \|_\infty$ for some $q_\infty > 0$.
%\begin{enumerate}
%	\item $\mathbf{E} [\compress(v)] = v$.
%	\item $\mathbf{E} \vert   \compress(v)_j - v_j\vert^2 \leq q_2 \vert v_j \vert^2$ for some $q_2 >0$.
%	\item $\| \compress(v) - v \|_\infty \leq q_\infty \| v \|_\infty$ for some $q_\infty > 0$. 
%\end{enumerate}		
\end{definition}	
\sparsifiedk~is one popular operator satisfying Definition~\ref{def:unbiasedcomp}. 
The \sparsifiedk~operator transmits on average $K$ elements of a vector $v\in\mathbb{R}^d$, which is defined by  
\begin{eqnarray}\label{eqn:sparsifiedK}
	[\compress(v)]_j = (v_j/p_j)\xi_j,
\end{eqnarray}
where $v_j$ is the $j^{\text{th}}$-element of $v\in\mathbb{R}^d$, $\xi_j\sim \texttt{Bernouli}(p_j)$ and $K=\sum_{j=1}^d p_j$.
The \sparsifiedk~operator captures several variants of sparsification in communication-efficient training. 
For example, \sparsifiedk~with $p_j = \vert v_j \vert/\| v \|_q$ and with $q=2$, $q=\infty$ and $q\in (0,\infty)$ is 1-bit QSGD in \cite{alistarh2017qsgd}, Terngrad in \cite{wen2017terngrad} and $\ell_q$-quantization in \cite{wang2018atomo}. Finally, the next lemma shows that \sparsifiedk~follows Definition~\ref{def:unbiasedcomp}.
\begin{lemma}\label{lemma:sparsifiedK}
	Let  $\compress(v)$ be \sparsifiedk~and $p_{\min} = \min_{j=[1,d]} p_j$. Then, $\compress(v)$ is \unbiasedcomp~with $q_2 = 1/p_{\min} -1$ and $q_\infty = \max(1/p_{\min} -1,1)$.
% for $j\in [1,d]$ and $v\in\mathbb{R}^d$
%	\begin{enumerate}
%		\item $\mathbf{E} [\compress(v)] = v$.
%		\item $\mathbf{E} \vert   \compress(v)_j - v_j\vert^2 \leq q_2 \vert v_j \vert^2$ with $q_2 = 1/p_{\min} -1$.
%		\item $\| \compress(v) - v \|_\infty \leq q_\infty \| v \|_\infty$ with $q_\infty = \max(1/p_{\min} -1,1)$. 
%	\end{enumerate}	
\end{lemma}

Next, we provide the definition of \biasedcomp, which captures biased compression operators of interest. 

\begin{definition}[\biasedcomp]\label{def:biasedcomp}
	$\compress(v):\mathbb{R}^d\rightarrow\mathbb{R}^d$ is \biasedcomp~if it satisfies $\| \compress(v) - v \|_\infty \leq (1-\alpha) \| v \|_\infty$ for some $\alpha \in (0,1]$ and for all $v\in\mathbb{R}^d$.
\end{definition}	
This definition requires the contraction property in the $\ell_\infty$-norm, rather than the $\ell_2$-norm in existing works, e.g.  \cite{alistarh2018convergence,khirirat2020compressed,stich2018sparsified,mitra2023temporal}.
One example that satisfies Definition~\ref{def:biasedcomp} is \topk.   Instead of transmitting on average $K$ elements, \topk~sends $K$ elements with the highest absolute magnitude from a vector $v\in\mathbb{R}^d$. This can be done by applying the following operator 
\begin{eqnarray}
	[\compress(v)]_i = \begin{cases}
		v_i & \text{if } i \in I_K(v) \\ 
		0   & \text{otherwise} 
	\end{cases}
\end{eqnarray}
Here, $I_K(v)$ is the index set for $K$ largest components of $v$ in the absolute magnitude. 
Finally, the \topk ~operator achieves the contraction in the $\ell_\infty$-norm in the next lemma.
\begin{lemma}\label{lemma:topK}
	Let $\compress(v)$ be 	\topk~and $\max_{j\in I_K^c(v)} \vert v_j \vert < \| v \|_\infty$ where  $I_K^c(v) = [1,d] - I_K(v)$. Then, 
 $\compress(v)$ is \biasedcomp~with $\alpha =1 - {\max_{j\in I_K^c(v)} \vert v_j \vert}/{\| v \|_\infty}$.
 %for  $v\in\mathbb{R}^d$, $\| \compress(v) - v \|_\infty \leq (1-\alpha) \| v \|_\infty$ with $\alpha =1 - {\max_{j\in I_K^c(v)} \vert v_j \vert}/{\| v \|_\infty}$.
\end{lemma}

\section{Convergence Analysis}\label{sec:convergence}

To this end, we provide the non-asymptotic convergence of \compfedrl ~with \unbiasedcomp~and \biasedcomp. 

\subsection{\compfedrl~with \unbiasedcomp}

The first theorem presents the convergence for \compfedrl~with \unbiasedcomp.
%
%\subsection{Convergence Analysis of \compfedrl ~with \unbiasedcomp} \label{unbiasedcomp}
\begin{theorem}\label{thm:DirectComp}
Consider Algorithm \ref{alg:EFFedQSync} with \unbiasedcomp. Let $\beta,\eta \in (0,1]$ and $0 \leq \| Q_0 \|_\infty \leq \frac{1}{1-\gamma}$. 
%
%$\compress(v)$ satisfy (A) $\mathbf{E} [\compress(v)] = v$, (B) $\mathbf{E} \vert   \compress(v)_j - v_j\vert^2 \leq q_2 \vert v_j \vert^2$ for $q_2>0$ and for all $j=[1,d]$, and (C) $\| \compress(v) - v \|_\infty \leq q_\infty \| v \|_\infty$ for $q_\infty >0$. 
%
Then, with probability at least $1-\delta$
\begin{eqnarray*}
		\| \overline Q_T - Q^\star  \|_\infty 
		&\leq& \rho^T \| \overline Q_0 - Q^\star \|_\infty +  \sqrt{\frac{\eta}{I}} e_1 + \frac{2\gamma}{C} + \frac{1}{\sqrt{I} }e_2,
	\end{eqnarray*}	
	where $\rho=1-\beta + \beta(1-\eta)^K$, $e_1 = \frac{4\gamma}{C}  \sqrt{ \log \frac{4\vert \mathcal{S}\vert \vert \mathcal{A}\vert T K}{\delta} }\left( 1+ \frac{1}{\sqrt{\eta I}}  \sqrt{ \log \frac{4\vert \mathcal{S}\vert \vert \mathcal{A}\vert T K }{\delta}} \right)$, $e_2 = \frac{1}{(1-(1-\eta)^K)}\left( \sqrt{16 \frac{4  q_2 \vert \mathcal{S} \vert \vert \mathcal{A}\vert}{(1-\gamma)^2} \log \frac{4T}{\delta}  } + \frac{4}{3} \frac{2q_\infty \sqrt{I}}{1-\gamma} \log \frac{4T}{\delta} \right)$, and $C = (1-\gamma)(1-(1-\eta)^K)$.
\end{theorem}	
\paragraph{Discussions on Theorem~\ref{thm:DirectComp}:}
Theorem~\ref{thm:DirectComp} establishes non-asymptotic high-probability convergence with the linear rate depending on federated parameters $\beta,\eta, K$ and three residual error terms for \compfedrl~with \unbiasedcomp.
We make the following observations. 
First, there is a trade-off between convergence speed and solution accuracy when we adjust the learning rate $\eta$.
In particular, increasing $\eta$ improves the convergence speed while increasing the first residual error term. 
Second, the first and last error term depends on $1/\sqrt{I}$, thus implying that \compfedrl~with \unbiasedcomp~attains the speed-up with respect to the number of agents like multi-agent TD(0) learning in \cite{mitra2023temporal}.
Third, the second error term is insignificant when the discount factor $\gamma$ is small. 
Fourth, Theorem~\ref{thm:DirectComp} with $q_2=q_\infty=0$ provides the convergence for full-precision federated $Q$-learning.

We obtain the convergence for \compfedrl~with \unbiasedcomp~and $K=1$ in the next corollary: 
\begin{corollary}\label{corr:DirectComp}
	Consider Algorithm \ref{alg:EFFedQSync} with  \unbiasedcomp~and $K=1$ under the same settings as Theorem~\ref{thm:DirectComp}. Then, with probability at least $1-\delta$ 
	\begin{eqnarray*}
		\| \overline Q_T - Q^\star  \|_\infty 
		&\leq& (1-\beta\eta)^T \| \overline Q_0 - Q^\star \|_\infty +  \sqrt{\frac{1}{\eta I}} \bar e_1 + \frac{2\gamma}{\eta(1-\gamma)} + \frac{1}{\eta \sqrt{I} } \bar e_2,
	\end{eqnarray*}	
 where $\bar e_1 = \frac{4\gamma}{(1-\gamma)}  \sqrt{ \log \frac{4\vert \mathcal{S}\vert \vert \mathcal{A}\vert T}{\delta} }\left( 1+ \frac{1}{\sqrt{\eta I}}  \sqrt{ \log \frac{4\vert \mathcal{S}\vert \vert \mathcal{A}\vert T}{\delta}} \right)$, and \\ $\bar e_2 = \left( \sqrt{16 \frac{4  q_2 \vert  \mathcal{S}\vert \vert \mathcal{A}\vert}{(1-\gamma)^2} \log \frac{4T}{\delta}  } + \frac{4}{3} \frac{2q_\infty \sqrt{I}}{1-\gamma} \log \frac{4T}{\delta} \right)$.
\end{corollary}	
From Corollary~\ref{corr:DirectComp}, we obtain the lowest convergence bound by letting $\eta=1$, which yields:
	\begin{eqnarray*}
	\| \overline Q_T - Q^\star  \|_\infty 
	&\leq& (1-\beta)^T \| \overline Q_0 - Q^\star \|_\infty +  \sqrt{\frac{1}{ I}} \bar e_1 + \frac{2\gamma}{(1-\gamma)} + \frac{1}{ \sqrt{I} } \bar e_2.
\end{eqnarray*}	
From this bound, \compfedrl ~with $K=1$ and \unbiasedcomp ~close to the identity operator (i.e. $q_2\rightarrow 0$ and $q_\infty\rightarrow 0$) and with $I \geq \max \left( B_1, B_2 \right)$ where $B_1 = \left( \frac{4\gamma}{1-\gamma}\right)^2\frac{16 \log (4 \vert \mathcal{S} \vert \vert \mathcal{A} \vert T/ \delta)}{\epsilon^2}$ and $B_2 = \left( \frac{4\gamma}{1-\gamma}\right) \frac{4 \sqrt{\log (4 \vert \mathcal{S} \vert \vert \mathcal{A} \vert T/ \delta)} }{\epsilon}$ achieves $\| \overline Q_T - Q^\star\|_\infty \leq \epsilon$ within $T = \frac{1}{\beta}\log(4 \| \overline Q_0 - Q^\star\|_\infty / \epsilon)$ iterations, 
when $\gamma \leq \frac{\epsilon}{8+\epsilon}$.
If $\beta = (1-\gamma)^4\min(\epsilon,\epsilon^2)/\log(4 \| \overline Q_0 - Q^\star\|_\infty / \epsilon)$, then we have the  $\mathcal{O}\left( \frac{1}{(1-\gamma)^4 \min(\epsilon,\epsilon^2)} \right)$ complexity that is close to the lower bound for a single-agent $Q$-learning by \cite{li2024q}.

%
%When we let $\beta = (1-\gamma)^5 \epsilon^2$, we achieve the same $\mathcal{O}(\frac{1}{(1-\gamma)^5\epsilon^2} )$ sample complexity as FedSynQ in Theorem 2 of \cite{Blessing}.

\subsection{\compfedrl ~with \biasedcomp} \label{biasedcomp}
The next theorem shows the convergence for \compfedrl~with \biasedcomp. 
\begin{theorem}\label{thm:Errorfeedback}
Consider Algorithm \ref{alg:EFFedQSync} with  \biasedcomp. Let $\beta,\eta \in (0,1]$ and $0 \leq \| Q_0 \|_\infty \leq \frac{1}{1-\gamma}$. 
%Further assume $\|  \compress(v) - v\|_\infty \leq (1-\alpha)\| v \|_\infty$ for some $\alpha \in (0,1]$ and $v\in\mathbb{R}^d$. 
%
Then, with probability at least $1-\delta$
\begin{eqnarray*}
	\| \overline Q_T - Q^\star  \|_\infty 
    &\leq&  \rho^T \| \overline Q_0 - Q^\star \|_\infty   + \frac{4}{C}  \sqrt{\frac{\eta}{I} \log \frac{2\vert \mathcal{S}\vert \vert \mathcal{A}\vert T K}{\delta} } e_1   + \frac{2\gamma}{C} + \frac{2\beta (1-\alpha)}{\alpha(1-\gamma)} D,
\end{eqnarray*}	
where $\rho=1-\beta + \beta(1-\eta)^K$, $e_1= 1+ \frac{1}{\sqrt{\eta I}}  \sqrt{ \log \frac{2\vert \mathcal{S}\vert \vert \mathcal{A}\vert T K }{\delta}}$, $C = (1-\gamma)(1-(1-\eta)^K)$, and $D= 1 + \frac{1+(1-\eta)^K}{1-(1-\eta)^K}$.
	
\end{theorem}	
\paragraph{Discussions on Theorem~\ref{thm:Errorfeedback}:}
From Theorem~\ref{thm:Errorfeedback} \compfedrl~with \biasedcomp~converges linearly with high probability with three residual error terms. 
The following observations from this theorem are similar to those from Theorem~\ref{thm:DirectComp}.
First, there is a trade-off between convergence speed and solution accuracy of the algorithm when we adjust the learning rate $\eta$, the federated parameter $\beta$, and the number of local epochs $K$. 
In particular, the algorithm with $K \gg 1$ and $\beta,\eta\rightarrow 1$ achieves high speed at the price of low solution accuracy. 
Second, due to the dependency of the first error term in the convergence bound on $1/\sqrt{I}$, \compfedrl~with \biasedcomp~achieves the speedup with respect to the number of agents similarly to \compfedrl~with \unbiasedcomp~in Theorem~\ref{thm:DirectComp} and to multi-agent TD(0) learning using error-feedback analyzed by \cite{mitra2023temporal}.
Third, the last error term is due to the compression $\alpha$ and can be diminished by decreasing $\beta$ and increasing $\eta, K$, whereas the second error term is negligible when the discount factor $\gamma$ is close to zero.  
Fourth, we obtain the convergence for full-precision synchronous $Q$-learning when we let $\alpha=1$ in Theorem~\ref{thm:Errorfeedback}.

The next corollary presents the convergence of \compfedrl~with \biasedcomp~when we let the number of local epoch $K=1$: 
\begin{corollary}\label{cor:errorfeedback}
Consider Algorithm \ref{alg:EFFedQSync} with  \biasedcomp~and $K=1$ under the same settings as Theorem~\ref{thm:Errorfeedback}. Then, with probability at least $1-\delta$
\begin{eqnarray*}
	\| \overline Q_T - Q^\star  \|_\infty 
    &\leq&  (1-\beta\eta)^T \| \overline Q_0 - Q^\star \|_\infty   + \frac{4}{1-\gamma}  \sqrt{\frac{1}{\eta I} \log \frac{2\vert \mathcal{S}\vert \vert \mathcal{A}\vert T}{\delta} } e_1   + \frac{2\gamma}{\eta (1-\gamma)}.
\end{eqnarray*}	
%where $e_1= 1+ \frac{1}{\sqrt{\eta I}}  \sqrt{ \log \frac{2\vert \mathcal{S}\vert \vert \mathcal{A}\vert T }{\delta}}$.
\end{corollary}
From Corollary~\ref{cor:errorfeedback}, \compfedrl~with $K=1$ and \biasedcomp, unlike \unbiasedcomp, does not have the residual error term due to compression. 
%
%Hence, we have the following observations. First, decreasing $\beta$ diminishes the last two error terms due to the compression $\alpha$ while reducing the convergence speed. 
%
%Second, the first error term which depends on $1/\sqrt{I}$ can be diminished when the number of agents is large, while the second error term is small when $\gamma \rightarrow 0$.
%
Furthermore, \compfedrl~with $K=1$ and \biasedcomp~ attains the lowest convergence bound when we let $\eta=1$, which yields: 
\begin{eqnarray*}
	\| \overline Q_T - Q^\star  \|_\infty 
    &\leq&  (1-\beta)^T \| \overline Q_0 - Q^\star \|_\infty   + \frac{4}{1-\gamma}  \sqrt{\frac{1}{ I} \log \frac{2\vert \mathcal{S}\vert \vert \mathcal{A}\vert T}{\delta} } e_1   + \frac{2\gamma}{ 1-\gamma}.
\end{eqnarray*}	
In conclusion, this algorithm reaches $\|\overline Q_T -Q^\star \|_\infty \leq \epsilon$ within $T = \frac{1}{\beta}\log(3 \| \overline Q_0 - Q^\star\|/\epsilon)$ iterations when $I = \frac{16}{(1-\gamma)^2} \log \frac{2 \vert \mathcal{S} \vert \vert \mathcal{A} \vert T}{\delta}$ and $\gamma \leq \frac{\epsilon}{6+\epsilon}$. 
If $\beta = (1-\gamma)^4\min(\epsilon,\epsilon^2)/ \log(3 \| \overline Q_0 - Q^\star\|/\epsilon)$, then we achieve the  $\mathcal{O}\left( \frac{1}{(1-\gamma)^4\min(\epsilon,\epsilon^2)} \right)$ complexity which nearly matches the lower bound for the single-agent $Q$-learning by \cite{li2024q}.

\section{Experiments} \label{Experiments}
In this section, we carry out numerical experiments to showcase the effectiveness of the \compfedrl ~algorithm.
\subsection{Setup}
%\paragraph{Setup.} 
We examine five variations of the noisy grid-world environment, some of which contain walls (as illustrated in Figure \ref{Maps} of Appendix).
%5 \ref{Maps}
Details regarding these environments and more results can be found in Appendix \ref{apendix:env} and Appendix \ref{apendix:result}, respectively.
%C.1 \ref{apendix:env} 
%C.2 \ref{apendix:result}
The $Q$-function is initialized with zero entries, and we set $\gamma=0.8$. After each round of communication  $t$, we evaluate the estimated $Q$ through the root mean square error (RMSE) between the estimated $\overline{Q}_t$ and the optimal $Q^*$.

To measure the communication efficiency of compression, we not only monitor the RMSE in communication rounds but also track the number of bits used by each agent.
We evaluate communicated bits by considering floating point precision (FPP) to be 32 bits. 
%for single precision floating-points.
Then, in each round of communication, for FedRL (i.e., \compfedrl ~without compression), \compfedrl ~with \topk, and \compfedrl ~with \sparsifiedk, the payload consumes 
$ |S||A| \times FPP $ bits,
$ K \times \left(\lceil \log_2(|S||A|) \rceil + FPP\right) $ bits, and
$ K' \times \left(\lceil \log_2(|S||A|) \rceil + FPP\right)$ bits, respectively.
Here, $K'$ is the size of the selected set by \sparsifiedk ~to be sent, and the factor $\lceil \log_2(|S||A|) \rceil$ comes from indicating $K$ ($K'$) indices in the dimension of the state-action space.

\begin{figure*}[t] %[H]
\centering
  \includegraphics[width=0.49\textwidth]{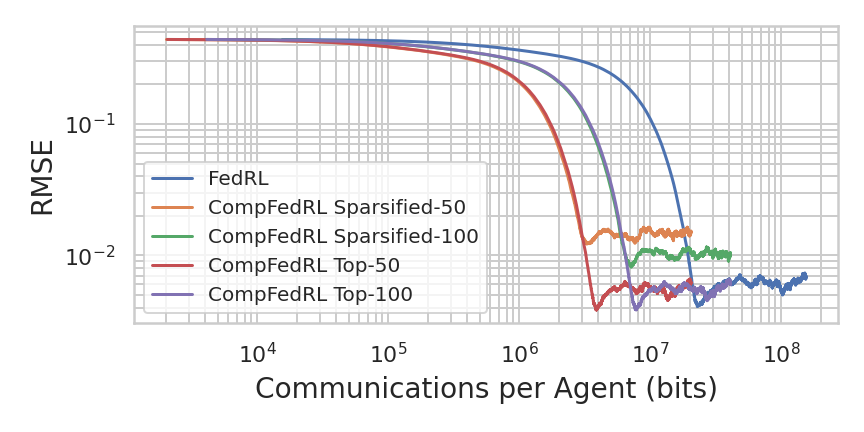}
  \includegraphics[width=0.49\textwidth]{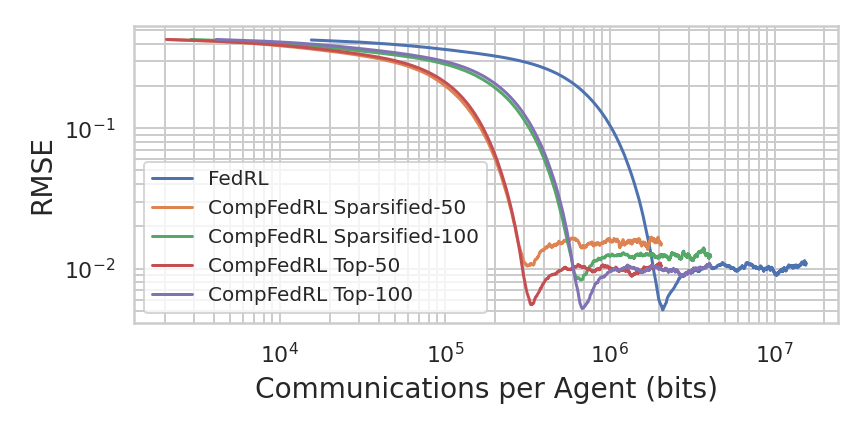}
\caption{Impact of compression: RMSE of the $Q$-estimates for both \textsf{FedRL} \cite{woo2023blessing, jin2022federated} (without compression) and \compfedrl ~to solve Map$11\times11$ task. Here, $I= 50$, $\eta =0.01$, $\beta= 0.8$, and (\textbf{Left}) $K=1$; (\textbf{Right}) $K=10$.}
%\caption{Impact of compression: the RMSE of the $Q$-estimates with respect to the number of bits communicated per agent for both \textsf{FedRL} \cite{woo2023blessing, jin2022federated} (without compression) and \compfedrl ~under different levels of sparsification via \sparsifiedk ~and \topk ~to solve Map$11\times11$ task. Here, the number of agents $I= 50$, the learning rate $\eta =0.01$, the federated parameter $\beta= 0.8$, and the number of local epochs (\textbf{Left}) $K=1$; (\textbf{Right}) $K=10$.}
\label{compression_Map4}
\end{figure*}

\begin{figure*}[t] %[H]
\centering
\includegraphics[width=0.49\textwidth]{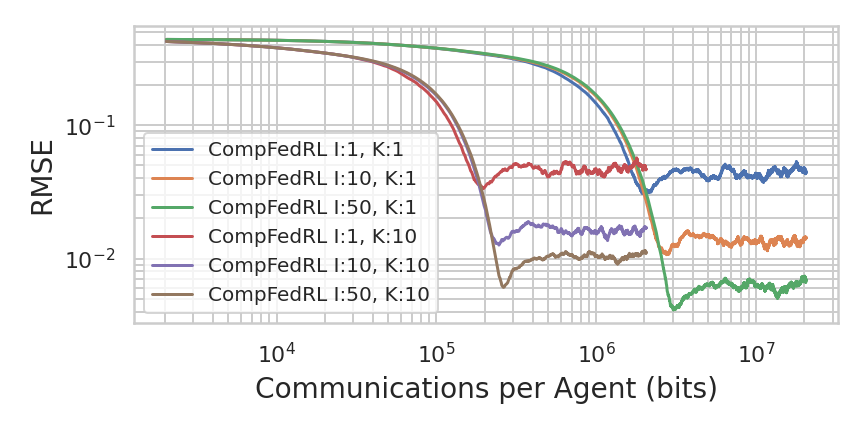}
\includegraphics[width=0.49\textwidth]{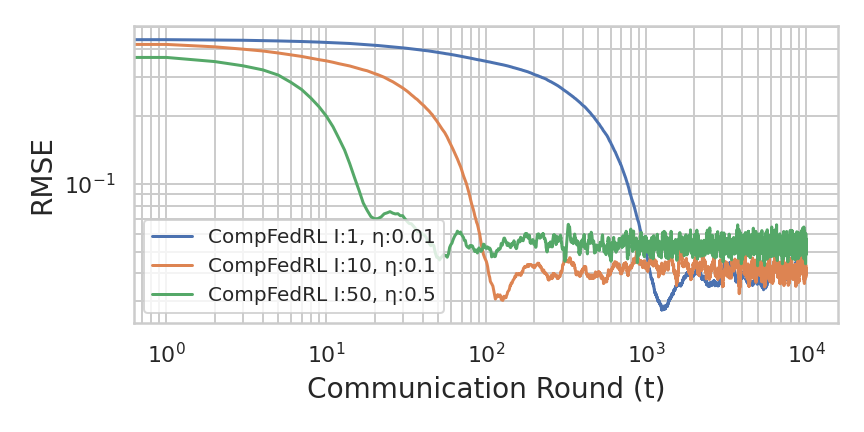}
\caption{Impact of the number of agents and local epochs: RMSE of the $Q$-estimates for \compfedrl ~under \textsf{Top-50} sparsification to solve Map$11\times11$ task. Here, (\textbf{Left}) $T\times K = 10000$, $\eta =0.01$, and $\beta=1$; (\textbf{Right}) $T= 10000$, $K=1$, and $\beta=0.8$.}
%\caption{Impact of the number of agents $I$ and the number of local epochs $K$: the RMSE of the $Q$-estimates for \compfedrl ~under \textsf{Top-50} sparsification to solve Map$11\times11$ task. Here, (\textbf{Left}) $T\times K = 10000$, the learning rate $\eta =0.01$, and the federated parameter $\beta=1$;(\textbf{Right}) the number of communication rounds $T= 10000$, the number of local epochs $K=1$, and the federated parameter $\beta=0.8$.}
\label{agent_K_speedup__Map4}
\end{figure*}

\begin{figure*}[t] %[H]
\centering
  \includegraphics[width=0.49\textwidth]{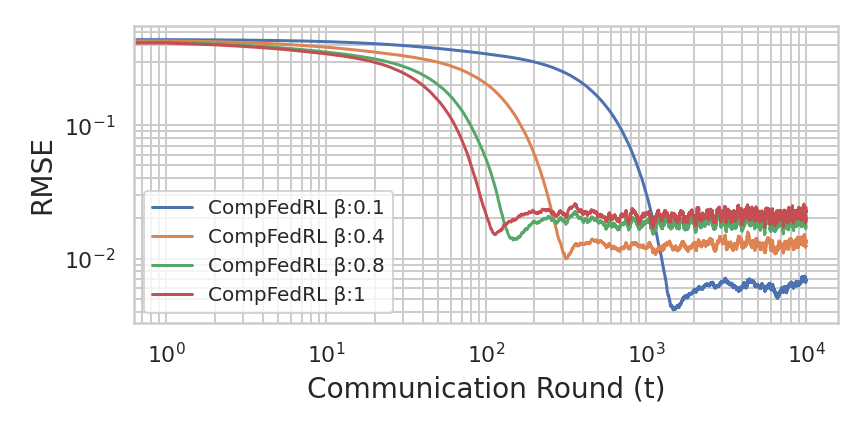}
  \includegraphics[width=0.49\textwidth]{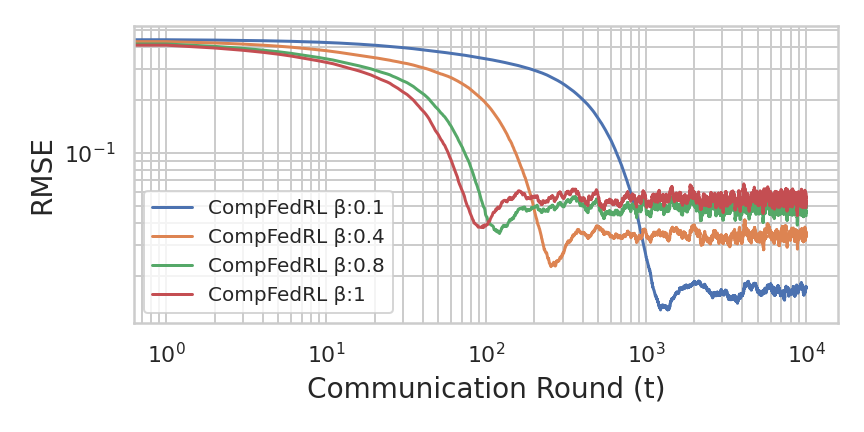}
\caption{Impact of federated parameter: RMSE of the $Q$-estimates for \compfedrl ~under (\textbf{Left}) \textsf{Top-50} ~and (\textbf{Right}) \textsf{Sparsified-50} ~sparsification to solve Map$11\times11$ task. Here, $I= 50$, $T= 10000$, $K=1$, and $\eta =0.1$.}
%\caption{Impact of federated parameter $\beta$: the RMSE of the $Q$-estimates with respect to the communication round for \compfedrl ~under (\textbf{Left}) \textsf{Top-50} ~and (\textbf{Right}) \textsf{Sparsified-50} ~sparsification to solve Map$11\times11$ task. Here, the number of agents $I= 50$, the number of communication rounds $T= 10000$, the number of local epochs $K=1$, and the learning rate $\eta =0.1$.}
\label{beta_Map4}
\end{figure*}

\begin{figure*}[t] %[H]
\centering
  \includegraphics[width=0.49\textwidth]{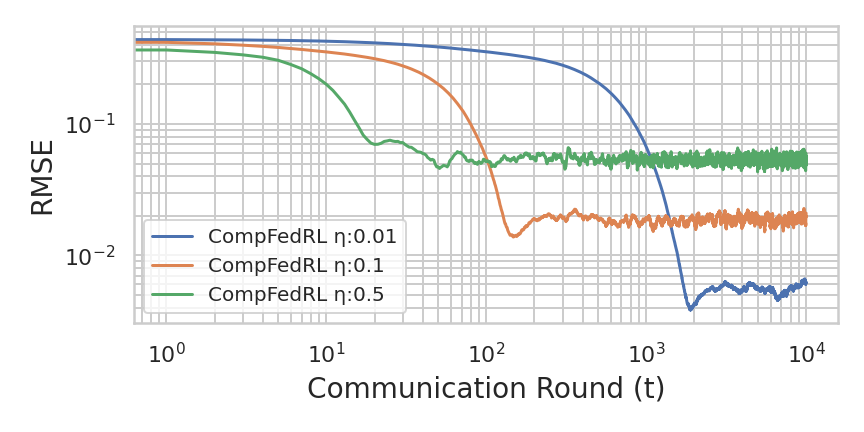}
  \includegraphics[width=0.49\textwidth]{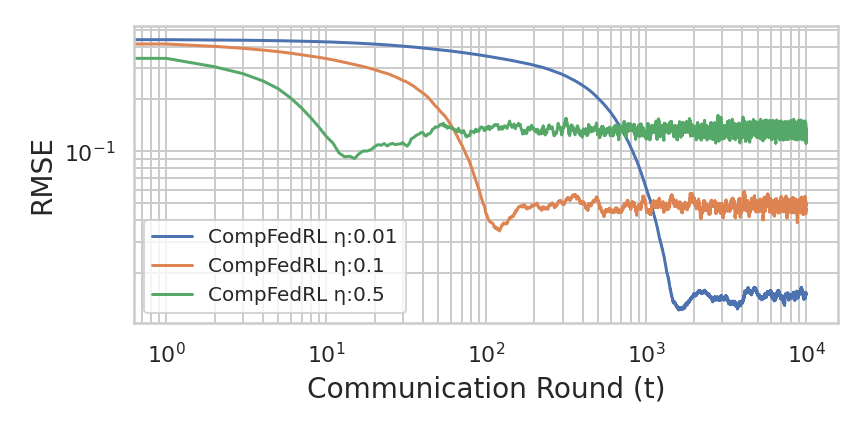}
\caption{Impact of learning rate: RMSE of the $Q$-estimates for \compfedrl ~under (\textbf{Left}) \textsf{Top-50} ~and (\textbf{Right}) \textsf{Sparsified-50} ~sparsification to solve Map$11\times11$ task. Here, $I= 50$, $T= 10000$, $K=1$, and $\beta=0.8$.}
%\caption{Impact of learning rate $\eta$: the RMSE of the $Q$-estimates with respect to the communication round for \compfedrl ~under (\textbf{Left}) \textsf{Top-50} ~and (\textbf{Right}) \textsf{Sparsified-50} ~sparsification to solve Map$11\times11$ task. Here, the number of agents $I= 50$, the number of communication rounds $T= 10000$, the number of local epochs $K=1$, and the federated parameter $\beta=0.8$.}
\label{eta_Map4}
\end{figure*}
\subsection{Communication Efficiency of \compfedrl}
%\paragraph{Communication Efficiency of \compfedrl.} 
Figure \ref{compression_Map4} demonstrates the impact of both \sparsifiedk ~and \topk~on reducing communication payload with the server and on convergence. 
Transitioning from local epochs of $K=1$ to $K=10$ showcases consistent convergence behavior despite a \emph{tenfold reduction} in communication frequency, underscoring the practicality of both \emph{periodic aggregation} and \emph{compression} mechanisms. 
A comparative analysis between \sparsifiedk ~and \topk ~reveals that \topk ~demonstrates lower residual error and converges to a point comparable to FedRL, as discussed in Section \ref{biasedcomp}, where it was highlighted that \biasedcomp ~does not incur a residual error term due to compression. Conversely, with \sparsifiedk, an increase in compression leads to a corresponding increase in residual error. 

\subsection{Convergence Speedup}
%\paragraph{Convergence speedup.} 
Local updating and the number of agents improve the performance of \compfedrl, as shown in Figure \ref{agent_K_speedup__Map4}. 
The left plot shows substantial communication reduction by local updating (e.g. $10$ times bits saving for $K=10$ against $K=1$) and confirms Theorem~\ref{thm:DirectComp} and~\ref{thm:Errorfeedback} that, given a fixed step size, an increase in the number of agents lowers residual error, improving convergence performance in RMSE. 
From the right plot, an interesting observation is that a higher number of agents allows for larger learning rates, which accelerates the convergence speed to attain almost the same target $\epsilon$-accuracy. 
%Additionally, an interesting observation is that to attain equivalent levels of $\epsilon$-accuracy, a higher agent count permits the utilization of larger learning rates, as evidenced by the graph on the right, facilitating a hastening of convergence.

\subsection{Impact of Federated Parameter $\beta$ and Learning Rate $\eta$}
%\paragraph{Impact of Federated Parameter $\beta$ and Learning Rate $\eta$}
%
Figure \ref{beta_Map4} and \ref{eta_Map4} demonstrate the trade-off between the convergence speed and solution accuracy when adjusting %the learning rate 
$\eta$ 
and %the federated parameter 
$\beta$. 
Increasing $\eta$ improves the speed while worsening the accuracy, as confirmed by Theorem~\ref{thm:DirectComp} and~\ref{thm:Errorfeedback} (see Figure~\ref{eta_Map4}).
Not captured by our theory, adjusting $\beta$ shows the same trade-off as $\eta$ as shown by Figure \ref{beta_Map4}.
Changes in $\eta$ affect residual error higher than adjustments in $\beta$, as supported by our non-asymptotic results emphasizing the dependency of most error terms on $\eta$.
Finally, thanks to its error-feedback mechanism, \topk ~shows greater robustness than \sparsifiedk ~to these parameter tunings. 

%Examining Figures \ref{beta_Map4} and \ref{eta_Map4}, a trade-off emerges between the convergence speed and solution accuracy of the algorithm when adjusting the learning rate $\eta$ and the federated parameter $\beta$, as discussed in Section \ref{biasedcomp}. However, a comparison between \topk ~and \sparsifiedk ~illustrates the greater robustness of \topk ~to variations in these parameters, owing to its error-feedback mechanism. Additionally, alterations in residual error due to changes in the learning rate $\eta$ exceed those resulting from adjustments in the federated parameter $\beta$. This observation aligns with our non-asymptotic results, where it is evident that most terms are contingent on $\eta$ rather than $\beta$.

\section{Conclusion}
We have proposed \compfedrl, a communication-efficient federated $Q$-learning algorithm that utilizes \emph{local epochs} and \emph{compression} to attain strong convergence performance while saving communication costs. 
Under \emph{synchronous} settings, we derive the non-asymptotic convergence of the algorithm using both \emph{unbiased} and \emph{biased} compression. 
Our theory suggests speed up with respect to the number of agents, and under certain choices of hyper-parameters \compfedrl ~attains the sample complexity that nearly matches the lower bound by \cite{li2024q}.
Experiments on the noisy grid-world environment corroborate our analyses and illustrate the benefits of periodically compressed updating and the number of agents on convergence performance of \compfedrl ~in terms of communication efficiency and solution accuracy. 
Our empirical results also affirm the speed-and-accuracy trade-off when adjusting the learning rates and federated parameters.

\section*{Acknowlegments}
This work was partially supported by the Swedish Research Council through grant agreement no. 2020-03607 and in part by Sweden's Innovation Agency (Vinnova). The computations were enabled by resources provided by the National Academic Infrastructure for Supercomputing in Sweden (NAISS) at Chalmers Centre for Computational Science and Engineering (C3SE) partially funded by the Swedish Research Council through grant agreement no. 2022-06725. 

\section*{Supplementary Material}
For detailed appendices and associated code, please refer to the full version of this paper available on Springer Link and arXiv \cite{beikmohammadi2024compressed}, as well as the code repository on GitHub: \url{https://github.com/AliBeikmohammadi/CompFedRL}.

%\bibliographystyle{splncs04}
%\bibliography{mybibliography}

\begin{thebibliography}{88}
\providecommand{\url}[1]{\texttt{#1}}
\providecommand{\urlprefix}{URL }
\providecommand{\doi}[1]{https://doi.org/#1}

\bibitem{alistarh2017qsgd}
Alistarh, D., Grubic, D., Li, J., Tomioka, R., Vojnovic, M.: {QSGD:} communication-efficient {SGD} via gradient quantization and encoding. Advances in neural information processing systems  \textbf{30} (2017)

\bibitem{alistarh2018convergence}
Alistarh, D., Hoefler, T., Johansson, M., Konstantinov, N., Khirirat, S., Renggli, C.: The convergence of sparsified gradient methods. Advances in Neural Information Processing Systems  \textbf{31} (2018)

\bibitem{assran2019gossip}
Assran, M., Romoff, J., Ballas, N., Pineau, J., Rabbat, M.: Gossip-based actor-learner architectures for deep reinforcement learning. Advances in Neural Information Processing Systems  \textbf{32} (2019)

\bibitem{azar2011speedy}
Azar, M.G., Munos, R., Ghavamzadeh, M., Kappen, H.: Speedy q-learning. In: Advances in neural information processing systems (2011)

\bibitem{beck2012error}
Beck, C.L., Srikant, R.: Error bounds for constant step-size q-learning. Systems \& control letters  \textbf{61}(12),  1203--1208 (2012)

\bibitem{beikmohammadi2024compressed}
Beikmohammadi, A., Khirirat, S., Magn{\'u}sson, S.: Compressed federated reinforcement learning with a generative model. arXiv preprint arXiv:2404.10635  (2024)

\bibitem{beikmohammadi2024distributed}
Beikmohammadi, A., Khirirat, S., Magn{\'u}sson, S.: Distributed momentum methods under biased gradient estimations. arXiv preprint arXiv:2403.00853  (2024)

\bibitem{beikmohammadi2024convergence}
Beikmohammadi, A., Khirirat, S., Magn{\'u}sson, S.: On the convergence of federated learning algorithms without data similarity. arXiv preprint arXiv:2403.02347  (2024)

\bibitem{beikmohammadi2023human}
Beikmohammadi, A., Magn{\'u}sson, S.: Human-inspired framework to accelerate reinforcement learning. arXiv preprint arXiv:2303.08115  (2023)

\bibitem{beikmohammadi2023ta}
Beikmohammadi, A., Magn{\'u}sson, S.: Ta-explore: Teacher-assisted exploration for facilitating fast reinforcement learning. In: Proceedings of the 2023 International Conference on Autonomous Agents and Multiagent Systems. pp. 2412--2414 (2023)

\bibitem{beikmohammadi2024accelerating}
Beikmohammadi, A., Magn{\'u}sson, S.: Accelerating actor-critic-based algorithms via pseudo-labels derived from prior knowledge. Information Sciences  \textbf{661},  120182 (2024)

\bibitem{bernstein2018signsgd}
Bernstein, J., Wang, Y.X., Azizzadenesheli, K., Anandkumar, A.: {signSGD}: Compressed optimisation for non-convex problems. In: International Conference on Machine Learning. pp. 560--569. PMLR (2018)

\bibitem{bertsekas1996neuro}
Bertsekas, D., Tsitsiklis, J.N.: Neuro-dynamic programming. Athena Scientific (1996)

\bibitem{beznosikov2023biased}
Beznosikov, A., Horv{\'a}th, S., Richt{\'a}rik, P., Safaryan, M.: On biased compression for distributed learning. Journal of Machine Learning Research  \textbf{24}(276),  1--50 (2023)

\bibitem{bhandari2018finite}
Bhandari, J., Russo, D., Singal, R.: A finite time analysis of temporal difference learning with linear function approximation. In: Conference on learning theory. pp. 1691--1692. PMLR (2018)

\bibitem{borkar2009stochastic}
Borkar, V.S.: Stochastic approximation: a dynamical systems viewpoint, vol.~48. Springer (2009)

\bibitem{borkar2000ode}
Borkar, V.S., Meyn, S.P.: The ode method for convergence of stochastic approximation and reinforcement learning. SIAM Journal on Control and Optimization  \textbf{38}(2),  447--469 (2000)

\bibitem{chen2020finite}
Chen, Z., Maguluri, S.T., Shakkottai, S., Shanmugam, K.: Finite-sample analysis of contractive stochastic approximation using smooth convex envelopes. Advances in Neural Information Processing Systems  \textbf{33},  8223--8234 (2020)

\bibitem{chen2021lyapunov}
Chen, Z., Maguluri, S.T., Shakkottai, S., Shanmugam, K.: A lyapunov theory for finite-sample guarantees of asynchronous q-learning and td-learning variants. arXiv preprint arXiv:2102.01567  (2021)

\bibitem{chen2022sample}
Chen, Z., Zhou, Y., Chen, R.R., Zou, S.: Sample and communication-efficient decentralized actor-critic algorithms with finite-time analysis. In: International Conference on Machine Learning. pp. 3794--3834. PMLR (2022)

\bibitem{chen2021multi}
Chen, Z., Zhou, Y., Chen, R.: Multi-agent off-policy td learning: Finite-time analysis with near-optimal sample complexity and communication complexity. arXiv preprint arXiv:2103.13147  (2021)

\bibitem{dal2023over}
Dal~Fabbro, N., Mitra, A., Heath, R., Schenato, L., Pappas, G.J.: Over-the-air federated td learning. Proceedings of the 6th MLSys Conference Workshop on Resource Constrained Learning in Wireless Networks  (2023)

\bibitem{dal2023federated}
Dal~Fabbro, N., Mitra, A., Pappas, G.J.: Federated td learning over finite-rate erasure channels: Linear speedup under markovian sampling. IEEE Control Systems Letters  (2023)

\bibitem{dalal2018finite}
Dalal, G., Sz{\"o}r{\'e}nyi, B., Thoppe, G., Mannor, S.: Finite sample analyses for td (0) with function approximation. In: Proceedings of the AAAI Conference on Artificial Intelligence. vol.~32 (2018)

\bibitem{doan2019finite}
Doan, T., Maguluri, S., Romberg, J.: Finite-time analysis of distributed td (0) with linear function approximation on multi-agent reinforcement learning. In: International Conference on Machine Learning. pp. 1626--1635. PMLR (2019)

\bibitem{doan2021finite}
Doan, T.T., Maguluri, S.T., Romberg, J.: Finite-time performance of distributed temporal-difference learning with linear function approximation. SIAM Journal on Mathematics of Data Science  \textbf{3},  298--320 (2021)

\bibitem{espeholt2018impala}
Espeholt, L., Soyer, H., Munos, R., Simonyan, K., Mnih, V., Ward, T., Doron, Y., Firoiu, V., Harley, T., Dunning, I., et~al.: Impala: Scalable distributed deep-rl with importance weighted actor-learner architectures. In: International conference on machine learning. pp. 1407--1416. PMLR (2018)

\bibitem{even2003learning}
Even-Dar, E., Mansour, Y., Bartlett, P.: Learning rates for q-learning. Journal of machine learning Research  \textbf{5}(1) (2003)

\bibitem{fan2021fault}
Fan, X., Ma, Y., Dai, Z., Jing, W., Tan, C., Low, B.K.H.: Fault-tolerant federated reinforcement learning with theoretical guarantee. Advances in Neural Information Processing Systems  \textbf{34},  1007--1021 (2021)

\bibitem{gheshlaghi2013minimax}
Gheshlaghi~Azar, M., Munos, R., Kappen, H.J.: Minimax pac bounds on the sample complexity of reinforcement learning with a generative model. Machine learning  \textbf{91},  325--349 (2013)

\bibitem{gorbunov2020linearly}
Gorbunov, E., Kovalev, D., Makarenko, D., Richt{\'a}rik, P.: Linearly converging error compensated sgd. Advances in Neural Information Processing Systems  \textbf{33},  20889--20900 (2020)

\bibitem{hu2019characterizing}
Hu, B., Syed, U.: Characterizing the exact behaviors of temporal difference learning algorithms using markov jump linear system theory. Advances in neural information processing systems  \textbf{32} (2019)

\bibitem{jaakkola1993convergence}
Jaakkola, T., Jordan, M., Singh, S.: Convergence of stochastic iterative dynamic programming algorithms. Advances in neural information processing systems  \textbf{6} (1993)

\bibitem{jin2022federated}
Jin, H., Peng, Y., Yang, W., Wang, S., Zhang, Z.: Federated reinforcement learning with environment heterogeneity. In: International Conference on Artificial Intelligence and Statistics. pp. 18--37. PMLR (2022)

\bibitem{karimireddy2019error}
Karimireddy, S.P., Rebjock, Q., Stich, S., Jaggi, M.: Error feedback fixes signsgd and other gradient compression schemes. In: International Conference on Machine Learning. pp. 3252--3261. PMLR (2019)

\bibitem{khirirat2020compressed}
Khirirat, S., Magn{\'u}sson, S., Johansson, M.: Compressed gradient methods with hessian-aided error compensation. IEEE Transactions on Signal Processing  \textbf{69},  998--1011 (2020)

\bibitem{khodadadian2022federated}
Khodadadian, S., Sharma, P., Joshi, G., Maguluri, S.T.: Federated reinforcement learning: Linear speedup under markovian sampling. In: International Conference on Machine Learning. pp. 10997--11057. PMLR (2022)

\bibitem{lakshminarayanan2018linear}
Lakshminarayanan, C., Szepesvari, C.: Linear stochastic approximation: How far does constant step-size and iterate averaging go? In: International Conference on Artificial Intelligence and Statistics. pp. 1347--1355. PMLR (2018)

\bibitem{li2024q}
Li, G., Cai, C., Chen, Y., Wei, Y., Chi, Y.: Is q-learning minimax optimal? a tight sample complexity analysis. Operations Research  \textbf{72}(1),  222--236 (2024)

\bibitem{li2020sample}
Li, G., Wei, Y., Chi, Y., Gu, Y., Chen, Y.: Sample complexity of asynchronous q-learning: Sharper analysis and variance reduction. Advances in neural information processing systems  \textbf{33},  7031--7043 (2020)

\bibitem{lin2022differentially}
Lin, C.Y., Kostina, V., Hassibi, B.: Differentially quantized gradient methods. IEEE Transactions on Information Theory  \textbf{68}(9),  6078--6097 (2022)

\bibitem{maei2018convergent}
Maei, H.R.: Convergent actor-critic algorithms under off-policy training and function approximation. arXiv preprint arXiv:1802.07842  (2018)

\bibitem{mandic2004generalized}
Mandic, D.P.: A generalized normalized gradient descent algorithm. IEEE signal processing letters  \textbf{11}(2),  115--118 (2004)

\bibitem{mayekar2020ratq}
Mayekar, P., Tyagi, H.: Ratq: A universal fixed-length quantizer for stochastic optimization. In: International Conference on Artificial Intelligence and Statistics. pp. 1399--1409. PMLR (2020)

\bibitem{mcmahan2017communication}
McMahan, B., Moore, E., Ramage, D., Hampson, S., y~Arcas, B.A.: Communication-efficient learning of deep networks from decentralized data. In: Artificial intelligence and statistics. pp. 1273--1282. PMLR (2017)

\bibitem{mitra2021linear}
Mitra, A., Jaafar, R., Pappas, G.J., Hassani, H.: Linear convergence in federated learning: Tackling client heterogeneity and sparse gradients. Advances in Neural Information Processing Systems  \textbf{34},  14606--14619 (2021)

\bibitem{mitra2023temporal}
Mitra, A., Pappas, G.J., Hassani, H.: Temporal difference learning with compressed updates: Error-feedback meets reinforcement learning. arXiv preprint arXiv:2301.00944  (2023)

\bibitem{mnih2016asynchronous}
Mnih, V., Badia, A.P., Mirza, M., Graves, A., Lillicrap, T., Harley, T., Silver, D., Kavukcuoglu, K.: Asynchronous methods for deep reinforcement learning. In: International conference on machine learning. pp. 1928--1937. PMLR (2016)

\bibitem{mnih2015}
Mnih, V., Kavukcuoglu, K., Silver, D., Rusu, A.A., Veness, J., Bellemare, M.G., Graves, A., Riedmiller, M., Fidjeland, A.K., Ostrovski, G., et~al.: Human-level control through deep reinforcement learning. nature  \textbf{518}(7540),  529--533 (2015)

\bibitem{puterman2014markov}
Puterman, M.L.: Markov decision processes: discrete stochastic dynamic programming. John Wiley \& Sons (2014)

\bibitem{qu2020finite}
Qu, G., Wierman, A.: Finite-time analysis of asynchronous stochastic approximation and $ q $-learning. In: Conference on Learning Theory. pp. 3185--3205. PMLR (2020)

\bibitem{shen2023towards}
Shen, H., Zhang, K., Hong, M., Chen, T.: Towards understanding asynchronous advantage actor-critic: Convergence and linear speedup. IEEE Transactions on Signal Processing  (2023)

\bibitem{srikant2019finite}
Srikant, R., Ying, L.: Finite-time error bounds for linear stochastic approximation andtd learning. In: Conference on Learning Theory. pp. 2803--2830. PMLR (2019)

\bibitem{stich2018sparsified}
Stich, S.U., Cordonnier, J.B., Jaggi, M.: Sparsified sgd with memory. Advances in neural information processing systems  \textbf{31} (2018)

\bibitem{sun2020finite}
Sun, J., Wang, G., Giannakis, G.B., Yang, Q., Yang, Z.: Finite-time analysis of decentralized temporal-difference learning with linear function approximation. In: International Conference on Artificial Intelligence and Statistics. pp. 4485--4495. PMLR (2020)

\bibitem{sutton2018reinforcement}
Sutton, R.S., Barto, A.G.: Reinforcement learning: An introduction. MIT press (2018)

\bibitem{szepesvari1997asymptotic}
Szepesv{\'a}ri, C.: The asymptotic convergence-rate of q-learning. Advances in neural information processing systems  \textbf{10} (1997)

\bibitem{tadic2001convergence}
Tadi{\'c}, V.: On the convergence of temporal-difference learning with linear function approximation. Machine learning  \textbf{42},  241--267 (2001)

\bibitem{tsitsiklis1996analysis}
Tsitsiklis, J., Van~Roy, B.: Analysis of temporal-diffference learning with function approximation. Advances in neural information processing systems  \textbf{9} (1996)

\bibitem{tsitsiklis1994asynchronous}
Tsitsiklis, J.N.: Asynchronous stochastic approximation and q-learning. Machine learning  \textbf{16},  185--202 (1994)

\bibitem{wai2020convergence}
Wai, H.T.: On the convergence of consensus algorithms with markovian noise and gradient bias. In: 2020 59th IEEE Conference on Decision and Control (CDC). pp. 4897--4902. IEEE (2020)

\bibitem{wainwright2019stochastic}
Wainwright, M.J.: Stochastic approximation with cone-contractive operators: Sharper $\ell_{\infty}$-bounds for $q$-learning. arXiv preprint arXiv:1905.06265  (2019)

\bibitem{wang2023federated}
Wang, H., Mitra, A., Hassani, H., Pappas, G.J., Anderson, J.: Federated temporal difference learning with linear function approximation under environmental heterogeneity. arXiv preprint arXiv:2302.02212  (2023)

\bibitem{wang2018atomo}
Wang, H., Sievert, S., Liu, S., Charles, Z., Papailiopoulos, D., Wright, S.: Atomo: Communication-efficient learning via atomic sparsification. Advances in neural information processing systems  \textbf{31} (2018)

\bibitem{rec1}
Wang, L., Zhang, W., He, X., Zha, H.: Supervised reinforcement learning with recurrent neural network for dynamic treatment recommendation. In: Proceedings of the 24th ACM SIGKDD International Conference on Knowledge Discovery \& Data Mining. p. 2447–2456. KDD '18, Association for Computing Machinery, New York, NY, USA (2018). \doi{10.1145/3219819.3219961}, \url{https://doi.org/10.1145/3219819.3219961}

\bibitem{watkins1992q}
Watkins, C.J., Dayan, P.: Q-learning. Machine learning  \textbf{8},  279--292 (1992)

\bibitem{wen2017terngrad}
Wen, W., Xu, C., Yan, F., Wu, C., Wang, Y., Chen, Y., Li, H.: Terngrad: Ternary gradients to reduce communication in distributed deep learning. Advances in neural information processing systems  \textbf{30} (2017)

\bibitem{woo2023blessing}
Woo, J., Joshi, G., Chi, Y.: The blessing of heterogeneity in federated q-learning: Linear speedup and beyond. In: International Conference on Machine Learning. pp. 37157--37216. PMLR (2023)

%\bibitem{Blessing}
%Woo, J., Joshi, G., Chi, Y.: The blessing of heterogeneity in federated q-learning: Linear speedup and beyond. In: Proceedings of the 40th International Conference on Machine Learning. ICML'23, JMLR.org (2023)

\bibitem{wu2021byzantine}
Wu, Z., Shen, H., Chen, T., Ling, Q.: Byzantine-resilient decentralized policy evaluation with linear function approximation. IEEE Transactions on Signal Processing  \textbf{69},  3839--3853 (2021)

\bibitem{ref4}
Xu, Z., Liu, S., Wu, Z., Chen, X., Zeng, K., Zheng, K., Su, H.: PATROL: A Velocity Control Framework for Autonomous Vehicle via Spatial-Temporal Reinforcement Learning, p. 2271–2280. Association for Computing Machinery, New York, NY, USA (2021), \url{https://doi.org/10.1145/3459637.3482283}

\bibitem{zhang2024finite}
Zhang, C., Wang, H., Mitra, A., Anderson, J.: Finite-time analysis of on-policy heterogeneous federated reinforcement learning. arXiv preprint arXiv:2401.15273  (2024)

\bibitem{zhang2020provably}
Zhang, S., Liu, B., Yao, H., Whiteson, S.: Provably convergent two-timescale off-policy actor-critic with function approximation. In: International Conference on Machine Learning. pp. 11204--11213. PMLR (2020)

\bibitem{zheng2023federated}
Zheng, Z., Gao, F., Xue, L., Yang, J.: Federated q-learning: Linear regret speedup with low communication cost. arXiv preprint arXiv:2312.15023  (2023)

\end{thebibliography}

%\iffalse

\clearpage

\appendix 

%\sarit{Put Lemma 1 and 2 into A.1. Then, add one small subsection saying that $0 \leq \| \overline Q_0 \|_\infty \leq \frac{r_{\max}}{1-\gamma}$ to make the theory more general.}
%\ali{Note that there are typically two cases: first: $0 \leq r \leq r_{max}$, second $\vert r\vert \leq r_{max}.$}

\section{Proof of Main Results}
We present proofs for main lemmas and theorems in this section. 
\subsection{Useful lemmas}
We begin by stating useful lemmas for our analysis. 
The first lemma proves the convergence bound for a non-negative sequence $V_k$ that satisfies $V_{k+1} \leq \rho V_k +e$ for $e>0$ and $\rho\in (0,1)$.

\begin{lemma}\label{lemma:trick_sum_seq}
Let the non-negative sequence $\{V_k\}_{k \geq 0}$ satisfy $V_{k+1} \leq \rho V_k + e$. If $\rho\in(0,1)$ and $e>0$, then $V_k \leq \rho^k V_0 + e/(1-\rho)$.
\end{lemma}

The next lemma provides high-probability bounds using Freeman's inequality, which extends from Bernstein's inequality.

\begin{lemma}\label{lemma:trick_dist_Q_learning}
	Let $E = \frac{\eta}{I}\sum_{t=0}^{T-1} (1-\eta)^{T-1-t}\sum_{i=1}^I w^{(i)}_t$ where each $w^{(i)}_t \in \mathbb{R}$ is independent of each other, and satisfy $\mathbf{E}w^{(i)}_t=0$, $\| w^{(i)}_t \|_\infty = W$ and $\mathbf{E}\vert  w^{(i)}_t \vert^2 \leq \sigma^2$. Then, 
	\begin{eqnarray*}
		\left\vert  E \right\vert \leq \sqrt{ \frac{8 \eta   \sigma^2}{I} \log \frac{2}{\delta}} + \frac{4  W}{3I} \log \frac{2}{\delta}
	\end{eqnarray*}	
	with probability at least $1-\delta$. 
\end{lemma}	
\begin{proof}
	Define $z^{(i)}_t = \eta\sum_{t=0}^{T-1} (1-\eta)^{T-1-t} w_t^{(i)}$. Then, $E=(1/I)\sum_{i=1}^I z^{(i)}_t$ which satisfies the following properties 
	\begin{enumerate}
		\item $\|  z^{(i)}_t \|_\infty \leq  \| w^{(i)}_t \|_\infty \leq  W$, where the first inequality derives from the fact that $\sum_{t=0}^{T-1}(1-\eta)^{T-1-t} \leq \frac{1}{\eta}$. { Note that this step is different from the step for proving (82) in \cite{woo2023blessing}, which is incorrect because they must bound 
  $B_t(s,a)= \max \vert \sum_{k=1}^K z_i^k(s,a) \vert$, not $B_t(s,a)= \max \vert z_i^k(s,a)\vert$.}
		\item $\mathbf{E}[z^{(i)}_t]=0$, which comes from the fact that $\mathbf{E}w^{(i)}_t=0$. 
		\item Next, we have 
		\begin{eqnarray*}
			W_I & = &  \sum_{i=1}^{I} \mathbf{E} (z_t^{(i)})^2 \\
			& = &  \sum_{i=1}^{I}   \sum_{t=0}^{T-1} \eta^2(1-\eta)^{2(T-1-t)} \mathbf{E} (w_t^{(i)})^2 \\
			& \leq & I \sigma^2   \sum_{t=0}^{T-1} \eta^2(1-\eta)^{2(T-1-t)}  \leq  \eta I \sigma^2. 
		\end{eqnarray*}
	\end{enumerate}	
	By Theorem 4 of \cite{woo2023blessing} with $m=1$, we have 
	\begin{eqnarray*}
		\left\vert \sum_{i=1}^I  z_t^{(i)} \right\vert \leq \sqrt{8 \eta I  \sigma^2 \log \frac{2}{\delta}} + \frac{4  W}{3} \log \frac{2}{\delta}  
	\end{eqnarray*}	
	with probability at least $1-\delta$. This implies that with probability at least $1-\delta$, 
	\begin{eqnarray*}
		\left\vert  E \right\vert \leq \sqrt{ \frac{8 \eta   \sigma^2}{I} \log \frac{2}{\delta}} + \frac{4  W}{3I} \log \frac{2}{\delta}.
	\end{eqnarray*}

\end{proof}

By using Lemma~\ref{lemma:trick_dist_Q_learning}, we obtain a high-probability bound for $\left\| \frac{1}{I}\sum_{i=1}^I Q^{(i)}_{t,K} - Q^\star \right\|_\infty$ for Algorithm~\ref{alg:EFFedQSync} with both \unbiasedcomp~and \biasedcomp~as follows: 
\begin{lemma}\label{lemma:trick_federated}
	Consider Algorithm \ref{alg:EFFedQSync}. Then, with probability at least $1-\delta/T$ 
	\begin{eqnarray*}
		\left\| \frac{1}{I}\sum_{i=1}^I Q^{(i)}_{t,K} - Q^\star \right\|_\infty 
		& \leq & (1-\eta)^K\| \overline Q_t - Q^\star \|_\infty+ \frac{4\gamma}{1-\gamma}  \sqrt{\frac{\eta}{I} \log \frac{2\vert \mathcal{S}\vert \vert \mathcal{A}\vert T K}{\delta} } e_1 + \frac{2\gamma}{1-\gamma}, 
	\end{eqnarray*}	
	where $e_1= 1+ \frac{1}{\sqrt{\eta I}}  \sqrt{ \log \frac{2\vert \mathcal{S}\vert \vert \mathcal{A}\vert T K }{\delta}}$. 
\end{lemma}
\begin{proof}
	Define $\Delta_{t,k} = \frac{1}{I}\sum_{i=1}^I Q^{(i)}_{t,k} - Q^\star$. Then, 
	\begin{eqnarray*}
		\Delta_{t,k+1} = (1-\eta)\Delta_{t,k} + \frac{\eta}{I} \sum_{i=1}^I [r + \gamma P^{(i)}_{t,k+1} V^{(i)}_{t,k}- Q^\star], 
	\end{eqnarray*}		
	where the local transition matrix $P_{t,k+1}^{(i)}$ is $1$ if $s^\prime=s^{(i)}_{t,k+1}(s,a)$ and $0$ otherwise, and $V_{t,k}^{(i)}(s) = \max_{a^\prime\in\mathcal{A}}  Q^{(i)}_{t,k}(s,a^\prime)$.
	Since $Q^\star = r + \gamma P V^\star$, 
	\begin{eqnarray*}
		\Delta_{t,k+1} = (1-\eta)\Delta_{t,k}  + \frac{\gamma\eta}{I}\sum_{i=1}^I (P_{t,k+1}^{(i)} -P)V^{(i)}_{t,k} + \frac{\gamma\eta}{I} \sum_{i=1}^I P(V^{(i)}_{t,k}-V^\star).
	\end{eqnarray*}		
	From the definition of the $\ell_\infty$-norm and by the triangle inequality, 
	\begin{eqnarray*}
		\| \Delta_{t,K} \|_\infty 
		& \leq & (1-\eta)^K\| \Delta_{t,0} \|_\infty+ \| E^1_t \|_\infty + \| E^2_t\|_\infty, 
	\end{eqnarray*}	
	where $E^1_t= \frac{\gamma\eta}{I} \sum_{k=0}^{K-1} (1-\eta)^{K-1-k}\sum_{i=1}^I (P_{t,k+1}^{(i)} -P)V^{(i)}_{t,k}$ and $E^2_t = \frac{\gamma\eta}{I} \sum_{k=0}^{K-1} (1-\eta)^{K-1-k}\sum_{i=1}^I P(V^{(i)}_{t,k} -V^\star)$. 

	To complete the proof, we must bound $\| E^1_t \|_\infty$ and $\| E^2_t \|_\infty$. We first bound we bound $\| E^2_t \|_\infty$.   By the fact that  $\| P(s,a)\|_1 \leq 1$,  $\| V^{(i)}_{t,k}\|_\infty$, and $\| V^\star\|_\infty \leq \frac{1}{1-\gamma}$, 
	\begin{eqnarray}
		\| E^2_t \|_\infty 
		& \leq & \frac{\gamma\eta}{I} \sum_{k=0}^{K-1} (1-\eta)^{K-1-k}\sum_{i=1}^I \| P \|_1( \| V^{(i)}_{t,k} \|_\infty + \| V^\star\|_\infty)  \notag \\
		& \leq & \frac{2\gamma}{1-\gamma} \sum_{k=0}^{K-1} \eta (1-\eta)^{K-1-k} \leq \frac{2\gamma}{1-\gamma}. \label{eqn:E_2_dist}
	\end{eqnarray}	
	
	Next, we bound  $\| E^1_t \|_\infty$ by Lemma~\ref{lemma:trick_dist_Q_learning} with $w_k^{(i)}=(P_{t, k+1}^{(i)} -P)V^{(i)}_{t,k}$. By the fact that $w_k^{(i)}$ is independent of the transition events of other agents, and that $\| P^{(i)}_{t,k}(s,a)\|_1 \leq 1$, $\| P(s,a)\|_1 \leq 1$, and $\| V^{(i)}_{t,k}\|_\infty \leq \frac{1}{1-\gamma}$, we can prove by proof arguments in Lemma~1 of \cite{woo2023blessing} that $W=\frac{2}{1-\gamma}$ and $\sigma^2 = \frac{2}{(1-\gamma)^2}$. Therefore, for every $s,a,t$
	\begin{eqnarray}
		\| E^1_t \|_\infty 
		&\leq& \gamma \left( \sqrt{ \frac{16 \eta  }{I(1-\gamma)^2} \log \frac{2\vert \mathcal{S}\vert \vert \mathcal{A}\vert T}{\delta}} + \frac{8  }{3I(1-\gamma)} \log \frac{2\vert \mathcal{S}\vert \vert \mathcal{A}\vert T}{\delta} \right) \notag \\
		& \leq & \frac{4\gamma}{1-\gamma}  \sqrt{\frac{\eta}{I} \log \frac{2\vert \mathcal{S}\vert \vert \mathcal{A}\vert T}{\delta} } \left( 1+ \frac{1}{\sqrt{\eta I}}  \sqrt{ \log \frac{2\vert \mathcal{S}\vert \vert \mathcal{A}\vert T}{\delta}} \right) \label{eqn:E_1_dist}
	\end{eqnarray}	
	with probability at least $1- \frac{\delta}{\vert \mathcal{S}\vert \vert \mathcal{A}\vert T}$.

	%
	%\newpage 
	%
	%
	%
	%Next, we can bound $\| E^1_t \|_\infty$ and $\| E^2_t \|_\infty$, as in \eqref{eqn:E_1_dist} and \eqref{eqn:E_2_dist}, respectively. 
	%
	%For any $s,a,t,k$, 
	%\begin{eqnarray*}
	%	\| E_1^t \|_\infty 
	%	& \leq & \frac{4\gamma}{1-\gamma}  \sqrt{\frac{\eta}{I} \log \frac{2\vert \mathcal{S}\vert \vert \mathcal{A}\vert T K}{\delta} } \left( 1+ \frac{1}{\sqrt{\eta I}}  \sqrt{ \log \frac{2\vert \mathcal{S}\vert \vert \mathcal{A}\vert T K }{\delta}} \right) 
	%\end{eqnarray*}	
	%with probability at least $1-\frac{\delta}{\vert \mathcal{S}\vert \vert \mathcal{A}\vert T K}$, and also   $\| E_2^t \|_\infty \leq \frac{2\gamma}{1-\gamma}$. 
	%
	Finally, by the union of bounds, we have with probability at least $1-\delta/T$ 
	\begin{eqnarray*}
		\| \Delta_{t,K} \|_\infty 
		& \leq & (1-\eta)^K\| \Delta_{t,0} \|_\infty+ \frac{4\gamma}{1-\gamma}  \sqrt{\frac{\eta}{I} \log \frac{2\vert \mathcal{S}\vert \vert \mathcal{A}\vert T K}{\delta} } e_1 + \frac{2\gamma}{1-\gamma}, 
	\end{eqnarray*}	
	where $e_1= 1+ \frac{1}{\sqrt{\eta I}}  \sqrt{ \log \frac{2\vert \mathcal{S}\vert \vert \mathcal{A}\vert T K }{\delta}}$. 
	Since $\Delta_{t,0} = \overline Q_t - Q^\star$, we complete the proof. 
\end{proof}

Next, Lemma~\ref{lemma:sparsifiedK} shows that $\sparsifiedk$ is $\unbiasedcomp$, whereas Lemma~\ref{lemma:topK} implies that $\topk$ is $\biasedcomp$. We now provide proof of these lemmas omitted from the main text. 
%
%The next two lemmas characterize the compression of interest used in our analysis. 
%
\subsubsection{Proof of Lemma~\ref{lemma:sparsifiedK}}
	We obtain the first statement by using \eqref{eqn:sparsifiedK}. Next, we prove the second statement. Since $\mathbf{E} \compress(v)_j = v_j$ and
	\begin{eqnarray*}
		\mathbf{E} \vert  \compress(v)_j  \vert^2 = \frac{1}{p_i}v_i^2,
	\end{eqnarray*}	
	we have 
	\begin{eqnarray*}
		\mathbf{E} \vert   \compress(v)_j - v_j\vert^2
		& = & \mathbf{E} \vert  \compress(v)_j  \vert^2 - \vert v_j \vert^2 \\
		& = & (1/p_i - 1) \vert v_j \vert^2 \\
		& \leq & (1/p_{\min} - 1)  \vert v_j \vert^2,
	\end{eqnarray*}	
	where $p_{\min} = \min_{j=[1,d]} p_i$. Finally, we prove the last statement. From the definition of the $\ell_\infty$-norm, 
	\begin{eqnarray*}
		\| \compress(v) - v \|_\infty  
		&  = & \max_{j\in [1,d]} \vert  \compress(v)_j - v_j  \vert \\ 
		&  \leq &  \max_{j\in [1,d]} \max(1/p_j - 1 ,1) \vert v_j \vert  \\
		& \leq & \max(1/p_{\min} -1 ,1)  \max_{j\in [1,d]}  \vert v_j \vert  = q_\infty \| v \|_\infty. 
	\end{eqnarray*}	

\subsubsection{Proof of Lemma~\ref{lemma:topK}}
	From  the definition of the $\ell_\infty$-norm and by denoting $I_K^c(v) = [1,d] - I_K(v)$,
	\begin{eqnarray*}
		\|  \compress(v) - v \|_\infty
		& = & \max_{j\in I_K^c(v)} \vert v_j \vert \\
		& = & \frac{\max_{j\in I_K^c(v)} \vert v_j \vert}{\| v \|_\infty} \| v \|_\infty. 
	\end{eqnarray*}	
	If $\max_{j\in I_K^c(v)} \vert v_j \vert < \| v \|_\infty$, then $	\|  \compress(v) - v \|_\infty \leq (1-\alpha) \| v \|_\infty$ with $\alpha =1 - {\max_{j\in I_K^c(v)} \vert v_j \vert}/{\| v \|_\infty}$.

\vspace{0.5cm}

To this end, by using the results above, we prove the main theorems in this paper. 

%we prove Lemma~\ref{lemma:sparsifiedK} and~\ref{lemma:topK}, which characterize the properties of \sparsifiedk and \topk, respectively. 
%

%\subsection{Proofs for Theorem~\ref{thm:DirectComp} and~\ref{thm:Errorfeedback}}

\subsection{Proof of Theorem~\ref{thm:DirectComp}}
Define $\Delta_t = \overline Q_t - Q^\star$. Then, the update for Algorithm \ref{alg:EFFedQSync} with \unbiasedcomp~can be expressed equivalently as 
\begin{eqnarray*}
	\Delta_{t+1} = (1-\beta)\Delta_t + \beta \left( \frac{1}{I}\sum_{i=1}^I Q^{(i)}_{t,K} - Q^\star \right) + \beta \frac{1}{I}\sum_{i=1}^I e^{(i)}_t, 
\end{eqnarray*}	
where $e^{(i)}_t = \compress(Q_{t,K}^{(i)}-\overline Q_{t} ) - (Q_{t,K}^{(i)}-\overline Q_{t} )$.  
From the definition of the $\ell_\infty$-norm and by the triangle inequality, 
\begin{eqnarray*}
	\| \Delta_{t+1} \|_\infty \leq (1-\beta) \| \Delta_t \|_\infty + \beta \left\| \frac{1}{I}\sum_{i=1}^I Q^{(i)}_{t,K} - Q^\star \right\|_\infty + \beta \left\| \frac{1}{I}\sum_{i=1}^I e^{(i)}_t \right\|_\infty.  
\end{eqnarray*}	
To complete the proof, we must bound $\left\| \frac{1}{I}\sum_{i=1}^I e^{(i)}_t \right\|_\infty$. 
%
%Let $v_j$ be $j^{\text{th}}$-element of $v\in\mathbb{R}^d$, and let $\compress(v)$ satisfy (A) $\mathbf{E} [\compress(v)] = v$, (B) $\mathbf{E} \vert   \compress(v)_j - v_j\vert^2 \leq q_2 \vert v_j \vert^2$ for $q_2>0$ and for all $j=[1,d]$, and (C) $\| \compress(v) - v \|_\infty \leq q_\infty \| v \|_\infty$ for $q_\infty >0$. 
%
Since from Lemma 4 of \cite{li2024q}, $\| Q^{(i)}_{t,K} - \overline Q_t \|_\infty  \leq \| Q_{t,K}^{(i)} \|_\infty + \| \overline Q_t\|_\infty  \leq \frac{2}{1-\gamma}$. By Lemma~\ref{lemma:sparsifiedK} we can then prove that 
\begin{eqnarray*}
	\mathbf{E} e_t^{(i)} & =&  0 \\ 
	\| e_t^{(i)}  \|_\infty & \leq &  q_\infty \| Q^{(i)}_{t,K} - \overline Q_t  \|_\infty  \leq  \frac{2q_\infty}{1-\gamma} \\ 
	\sum_{i=1}^I \mathbf{E} (e_t^{(i)})^2 & \leq & \sum_{i=1}^I q_2 \| Q^{(i)}_{t,K} - \overline Q_t  \|_2^2 \leq \frac{4 I q_2 \vert \mathcal{S}\vert \vert  \mathcal{A}\vert}{(1-\gamma)^2}.
\end{eqnarray*}	
By using Theorem 4 of \cite{woo2023blessing} with $m=1$,  we have 
 \begin{eqnarray*}
  \left\| \frac{1}{I}\sum_{i=1}^I e^{(i)}_t \right\|_\infty \leq  \frac{1}{\sqrt{I}} \left( \sqrt{16 \frac{4  q_2 \vert  \mathcal{S} \vert \vert \mathcal{A}\vert}{(1-\gamma)^2} \log \frac{4T}{\delta}  } + \frac{4}{3} \frac{2q_\infty \sqrt{I}}{1-\gamma} \log \frac{4T}{\delta} \right). 
 \end{eqnarray*}	
 with probability $1-\delta/(2T)$. Plugging this above result and the result from  Lemma~\ref{lemma:trick_federated} with $\delta^\prime=\delta/2$ into the main inequality, and using the union of bounds, we have  
 \begin{eqnarray*}
 	\| \Delta_{T} \|_\infty 
 	&\leq& (1-\beta + \beta(1-\eta)^K) \| \Delta_t \|_\infty +  \frac{4\beta\gamma}{1-\gamma}  \sqrt{\frac{\eta}{I} \log \frac{4 \vert \mathcal{S}\vert \vert \mathcal{A}\vert T K}{\delta} } e_1  \\
 	&&  + \frac{2\beta\gamma}{1-\gamma} + \frac{\beta}{\sqrt{I}}e_2,
 \end{eqnarray*}	
 with probability at least $1-\delta/T$. Here, $e_1= 1+ \frac{1}{\sqrt{\eta I}}  \sqrt{ \log \frac{4\vert \mathcal{S}\vert \vert \mathcal{A}\vert T K }{\delta}}$ and $e_2 = \sqrt{16 \frac{4  q_2 \vert \mathcal{S} \vert \vert \mathcal{A}\vert}{(1-\gamma)^2} \log \frac{4T}{\delta}  } + \frac{4}{3} \frac{2q_\infty \sqrt{I}}{1-\gamma} \log \frac{4T}{\delta}$.
Finally from Lemma~\ref{lemma:trick_sum_seq},  with probability at least $1-\delta$
 \begin{eqnarray*}
	\| \Delta_{T} \|_\infty 
	&\leq& (1-\beta + \beta(1-\eta)^K)^T \| \Delta_0 \|_\infty +  \frac{4\gamma}{C}  \sqrt{\frac{\eta}{I} \log \frac{4\vert \mathcal{S}\vert \vert \mathcal{A}\vert T K}{\delta} } e_1  \\
	&&  + \frac{2\gamma}{C} + \frac{1}{\sqrt{I} (1-(1-\eta)^K)}e_2,
\end{eqnarray*}	
where $C = (1-\gamma)(1-(1-\eta)^K)$.

\subsection{Proof of Theorem~\ref{thm:Errorfeedback}}
Define $\hat \Delta_t = \overline Q_t + (\beta/I)\sum_{i=1}^I e^{(i)}_t- Q^\star$. Then, the update for Algorithm \ref{alg:EFFedQSync} with  \biasedcomp~can be expressed equivalently as 
\begin{eqnarray*}
	\hat \Delta_{t+1} = (1-\beta)\hat\Delta_t + \beta \left( \frac{1}{I}\sum_{i=1}^I Q^{(i)}_{t,K} - Q^\star \right) + \beta^2 \frac{1}{I}\sum_{i=1}^I e^{(i)}_t, 
\end{eqnarray*}	
where $e^{(i)}_{t+1} = Q_{t,K}^{(i)}-\overline Q_{t} + e^{(i)}_t - \compress(Q_{t,K}^{(i)}-\overline Q_{t} + e^{(i)}_t)$. From the definition of the $\ell_\infty$-norm and by the triangle inequality, 
\begin{eqnarray*}
	\| \hat \Delta_{t+1} \|_\infty  \leq  (1-\beta) \| \hat\Delta_t \|_\infty + \beta \left\| \frac{1}{I}\sum_{i=1}^I Q^{(i)}_{t,K} - Q^\star \right\|_\infty + \beta^2 \left\|  \frac{1}{I}\sum_{i=1}^I e^{(i)}_t \right\|_\infty. 
\end{eqnarray*}	
From Lemma~\ref{lemma:trick_federated}, with probability at least $1-\delta/T$ 
\begin{eqnarray*}
	\| \hat \Delta_{t+1} \|_\infty  &\leq&  (1-\beta) \| \hat\Delta_t \|_\infty + \beta (1-\eta)^K\| \overline Q_t - Q^\star \|_\infty \\
	&& + \frac{4\beta\gamma}{1-\gamma}  \sqrt{\frac{\eta}{I} \log \frac{2\vert \mathcal{S}\vert \vert \mathcal{A}\vert T K}{\delta} } e_1 + \frac{2\beta\gamma}{1-\gamma} + \beta^2 \left\|  \frac{1}{I}\sum_{i=1}^I e^{(i)}_t \right\|_\infty, 
\end{eqnarray*}	
where $e_1= 1+ \frac{1}{\sqrt{\eta I}}  \sqrt{ \log \frac{2\vert \mathcal{S}\vert \vert \mathcal{A}\vert T K }{\delta}}$. 
Since 
\begin{eqnarray*}
    \| \overline Q_t - Q^\star \|_\infty \leq \| \hat \Delta_t\|_\infty + \beta \left\| \frac{1}{I} \sum_{i=1}^I e^{(i)}_t \right\|_\infty,
\end{eqnarray*}
we get: with probability at least $1-\delta/T$ 
\begin{eqnarray*}
	\| \hat \Delta_{t+1} \|_\infty  &\leq&  (1-\beta + \beta(1-\eta)^K) \| \hat\Delta_t \|_\infty + \frac{4\beta\gamma}{1-\gamma}  \sqrt{\frac{\eta}{I} \log \frac{2\vert \mathcal{S}\vert \vert \mathcal{A}\vert T K}{\delta} } e_1   \\
	&& + \frac{2\beta\gamma}{1-\gamma} + \beta^2 (1 + (1-\eta)^K ) \left\|  \frac{1}{I}\sum_{i=1}^I e^{(i)}_t \right\|_\infty. 
\end{eqnarray*}	
Next, suppose $\|  \compress(v) - v\|_\infty \leq (1-\alpha)\| v \|_\infty$ for some $\alpha \in (0,1]$ and $v\in\mathbb{R}^d$. Define $w_t^{(i)}=e_t^{(i)}$. Since from Lemma 4 of \cite{li2024q}, $\| Q_{t,K}^{(i)} - \overline Q_t\|_\infty \leq \| Q_{t,K}^{(i)} \|_\infty + \| \overline Q_t\|_\infty \leq \frac{2}{1-\gamma}$. Therefore,  we have
\begin{eqnarray*}
	\|  e^{(i)}_{t+1}\|_\infty
	& \leq & (1-\alpha) \|  e^{(i)}_{t} + Q_{t,K}^{(i)} - \overline Q_t \|_\infty \\ 
	&\leq& (1-\alpha)\|  e^{(i)}_{t}\|_\infty + (1-\alpha)\| Q_{t,K}^{(i)} - \overline Q_t \|_\infty \\
	& \leq & (1-\alpha)\|  e^{(i)}_{t}\|_\infty + \frac{2(1-\alpha)}{1-\gamma}. 
\end{eqnarray*}	
From Lemma~\ref{lemma:trick_sum_seq} and by the fact that $e_0^{(i)}=0$, 
\begin{eqnarray*}
	\|  e^{(i)}_{t}\|_\infty \leq \frac{2(1-\alpha)}{\alpha(1-\gamma)}.
\end{eqnarray*}	
Thus, 
\begin{eqnarray*}
	\left\|  \frac{1}{I}\sum_{i=1}^I e^{(i)}_t \right\|_\infty
	& \leq &  \frac{1}{I}\sum_{i=1}^I	\|  e^{(i)}_{t}\|_\infty  \leq \frac{2(1-\alpha)}{\alpha(1-\gamma)}.
\end{eqnarray*}	
Plugging the upper-bound for $	\left\|  \frac{1}{I}\sum_{i=1}^I e^{(i)}_t \right\|_\infty$ into the main inequality yields 
\begin{eqnarray*}
	\| \hat \Delta_{t+1} \|_\infty  &\leq&  (1-\beta + \beta(1-\eta)^K) \| \hat\Delta_t \|_\infty + \frac{4\beta\gamma}{1-\gamma}  \sqrt{\frac{\eta}{I} \log \frac{2\vert \mathcal{S}\vert \vert \mathcal{A}\vert T K}{\delta} } e_1   \\
	&& + \frac{2\beta\gamma}{1-\gamma} + \beta^2 (1 + (1-\eta)^K ) \frac{2(1-\alpha)}{\alpha(1-\gamma)}
\end{eqnarray*}	
with probability at least $1-\delta/T$. Using Lemma~\ref{lemma:trick_sum_seq} and the union of bounds, we have the following: with probability at least $1-\delta$
\begin{eqnarray*}
	\| \hat \Delta_{T} \|_\infty  &\leq&  (1-\beta + \beta(1-\eta)^K)^T \| \hat\Delta_0 \|_\infty  \\
	&& + \frac{4}{(1-\gamma)(1-(1-\eta)^K)}  \sqrt{\frac{\eta}{I} \log \frac{2\vert \mathcal{S}\vert \vert \mathcal{A}\vert T K}{\delta} } e_1   \\
	&& + \frac{2\gamma}{(1-\gamma)(1-(1-\eta)^K)} + \beta (1 + (1-\eta)^K ) \frac{2(1-\alpha)}{\alpha(1-\gamma)(1-(1-\eta)^K)}. 
\end{eqnarray*}	
Since $e^{(i)}_0=0$, $\hat \Delta_0 = \Delta_0$.
Next, by the triangle inequality and by the upper-bound for $	\left\|  \frac{1}{I}\sum_{i=1}^I e^{(i)}_t \right\|_\infty$, we have with probability at least $1-\delta$
\begin{eqnarray*}
	\| \Delta_T \|_\infty 
	& \leq & \| \hat \Delta_{T} \|_\infty + \frac{\beta}{I}\sum_{i=1}^I \| e_t^{(i)}\|_\infty \\
	& \leq &  \| \hat \Delta_{T} \|_\infty + \frac{2\beta(1-\alpha)}{\alpha(1-\gamma)}  \\
	&\leq&  (1-\beta + \beta(1-\eta)^K)^T \| \Delta_0 \|_\infty  \\
	&& + \frac{4}{(1-\gamma)(1-(1-\eta)^K)}  \sqrt{\frac{\eta}{I} \log \frac{2\vert \mathcal{S}\vert \vert \mathcal{A}\vert T K}{\delta} } e_1   \\
	&& + \frac{2\gamma}{(1-\gamma)(1-(1-\eta)^K)} + \frac{2\beta (1-\alpha)}{\alpha(1-\gamma)} \left( 1 + \frac{1+(1-\eta)^K}{1-(1-\eta)^K} \right).
\end{eqnarray*}

\section{Discussion on the Boundedness of $\| \bar Q_0\|_\infty$}
We derive Theorem~\ref{thm:DirectComp} and~\ref{thm:Errorfeedback} under the assumptions that $\| r\|_\infty \leq 1$ and $\| \bar Q_0\|_\infty \leq \frac{1}{1-\gamma}$. 
These theorems, however, do not lose any generality, since they can be extended to the cases where $0 \leq r \leq r_{\max}$ or $\vert r \vert \leq r_{\max}$. This can be done by imposing $\| r\|_\infty \leq r_{\max}$ and $\| \bar Q_0\|_\infty \leq {r_{\max}}/{(1-\gamma)}$. These results imply that $\| Q^{(i)}_{t,K} - \bar Q_t \|_\infty \leq 2r_{\max}/(1-\gamma)$ and hence that the residual error terms in Theorem~\ref{thm:DirectComp} and~\ref{thm:Errorfeedback} contain $r_{\max}$ apart from other federated and problem parameters.

\section{Details of Experiments} \label{apendix:exp}

\subsection{Environments} \label{apendix:env}
\begin{figure*}[htp] %[H]
\centering
\subfloat[Map$5\times5$\label{Map1}]{
  \includegraphics[width=0.19\textwidth]{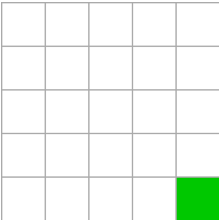}}
  \hfil
\subfloat[Map$5\times5$w\label{Map2}]{
  \includegraphics[width=0.19\textwidth]{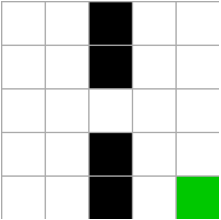}}
  \hfil
\subfloat[Map$6\times6$w\label{Map3}]{
  \includegraphics[width=0.19\textwidth]{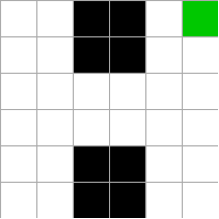}}
  \hfil
\subfloat[Map$11\times11$\label{Map4}]{
  \includegraphics[width=0.19\textwidth]{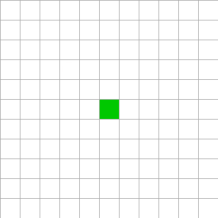}}
  \hfil
\subfloat[Map$17\times17$w\label{Map5}]{
  \includegraphics[width=0.19\textwidth]{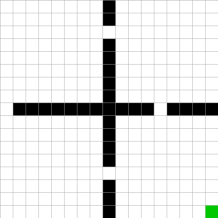}}
  \hfil
\caption{Various noisy Grid-world environments utilized as experimental tasks, requiring navigation through a grid towards a Goal (green tile) state while avoiding collision with Walls (black tiles) by traversing Empty (white tiles) cells.}
\label{Maps}
\end{figure*}

We use five noisy variants of the grid-world environment, a basic grid-based simulation often used in RL experiments, as shown in Figure \ref{Maps}. In this environment, agents navigate a grid composed of various cells, including walls, empty spaces, and a goal state. Actions available to the agent include moving up, down, left, or right. The state of the environment is represented by the current position of the agent within the grid. Rewards are defined based on the agent's interactions with the environment; reaching the goal state yields a positive reward $+1$, while taking a non-valid action, such as trying to move through a wall or exit the grid, incurs a penalty $-1$ and the agent stays in the same position.
To introduce noise into the environment, a clipped Gaussian noise model is employed, with the specified standard deviation and clipping threshold, set at 0.5 (i.e., $\textsf{clip}\{\mathcal{N}(\mu=0, \sigma^2=0.5), -0.5, 0.5\}$). This noise affects the agent's interaction, simulating real-world uncertainties and imperfections in the environment.

%\subsection{Baselines and Implementation Details}
%\subsection{Hyperparameters}
%\subsection{Ablation Study}

\subsection{Results and Discussion} \label{apendix:result}

\begin{figure*}[htp] %[H]
\centering
\subfloat[Map$5\times5$\label{p1}]{
  \includegraphics[width=0.185\textwidth]{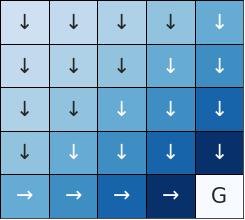}}
  \hfil
\subfloat[Map$5\times5$w\label{p2}]{
  \includegraphics[width=0.185\textwidth]{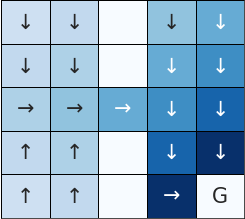}}
  \hfil
\subfloat[Map$6\times6$w\label{p3}]{
  \includegraphics[width=0.185\textwidth]{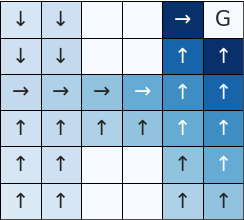}}
  \hfil
\subfloat[Map$11\times11$\label{p4}]{
  \includegraphics[width=0.185\textwidth]{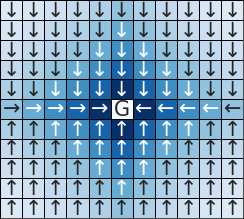}}
  \hfil
\subfloat[Map$17\times17$w\label{p5}]{
  \includegraphics[width=0.215\textwidth]{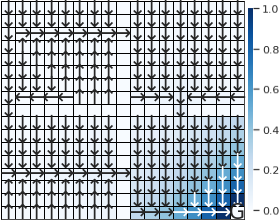}}
  \hfil
\caption{Learned $Q$-values and policy for different tasks. Arrows represent the action that learned policy proposes in each state. Here, agents have been trained using \compfedrl ~algorithm under (a-c )\textsf{Top-5} (d-e)\textsf{Top-50} ~sparsification with the number of agents $I= 50$, the number of communication rounds $T= 1000$, the number of local epochs $K=10$, the learning rate $\eta =0.01$, and the federated parameter $\beta=0.8$.}
\label{Policies}
\end{figure*}

Applying \compfedrl ~on our environments has been lead to finding the optimal policy in all cases. For example, see Figure \ref{Policies} where show $Q$-values in a heatmap with values correspond to colorbar. also arrows showing the action $a$ suggested with the policy learned by our algorithms. In this section, we all have shown the same bath of figures shown in the paper, but for the other four tasks. Results are fully in agreement with what we discussed in Section \ref{Experiments}, showing the applicability of algorithms and robustness to the range of task from easy to hard, from loss state-action space to high dimension. To seake of gravity, we omit reporting discussion here and refer the redere to Section \ref{Experiments}.

The application of \compfedrl ~to our task has consistently led to the discovery of optimal policies across all scenarios. For instance, Figure \ref{Policies} illustrates $Q$-values represented in a heatmap, with corresponding values indicated by the color bar, alongside arrows denoting the suggested action derived from the policy learned by our algorithms, showing the optimality of the founded policy. 

This section also presents a cohesive collection of figures akin to those showcased in the paper, each pertaining to one of the remaining four tasks. The outcomes are entirely congruent with the discussions presented in Section \ref{Experiments}, underscoring the versatility of the algorithms and their robustness across a spectrum of tasks ranging from straightforward to challenging, and from low-dimensional to high-dimensional state-action spaces. For brevity, we abstain from reiterating the discussions here and direct the reader to Section \ref{Experiments} for further details.

\subsubsection{Results for Map$5\times5$ task.} \label{apendix:Map1}
\begin{figure*}[!htb] %[H]
\centering
\subfloat[Number of local epochs $K=1$\label{compression_K1_Map1}]{
  \includegraphics[width=0.49\textwidth]{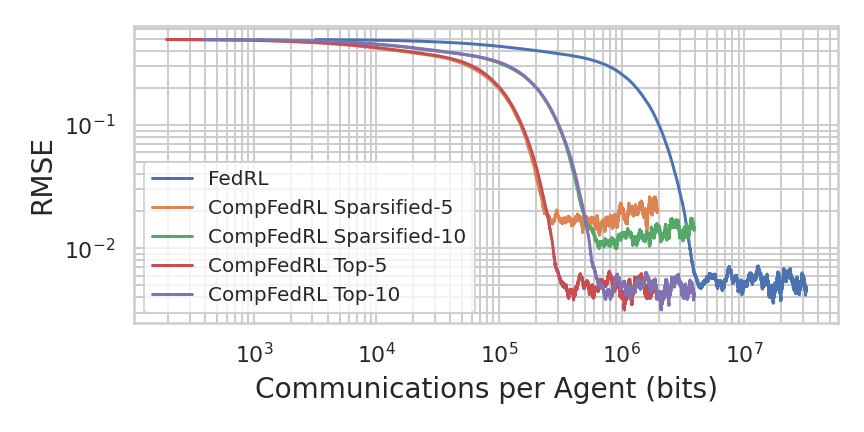}}
  \hfil
\subfloat[Number of local epochs $K=10$\label{compression_K10_Map1}]{
  \includegraphics[width=0.49\textwidth]{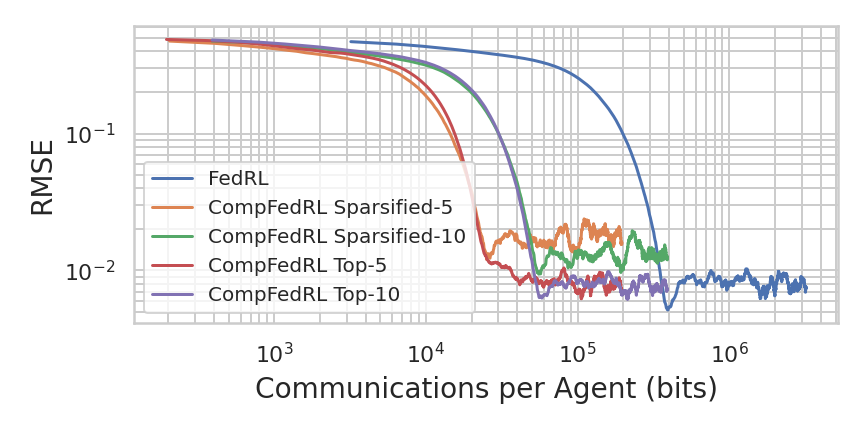}}
\caption{Impact of compression: the RMSE of the $Q$-estimates with respect to the number of bits communicated per agent for both \textsf{FedRL} (without compression) and \compfedrl ~under different levels of sparsification via \sparsifiedk ~and \topk ~to solve Map$5\times5$ task.
Here, the number of agents $I= 50$, the learning rate $\eta =0.01$, and the federated parameter $\beta= 0.8$.}
\label{compression_Map1}
\end{figure*}

\begin{figure*}[!htb] %[H]
\centering
\subfloat[\label{agent_K_Map1}]{
\includegraphics[width=0.49\textwidth]{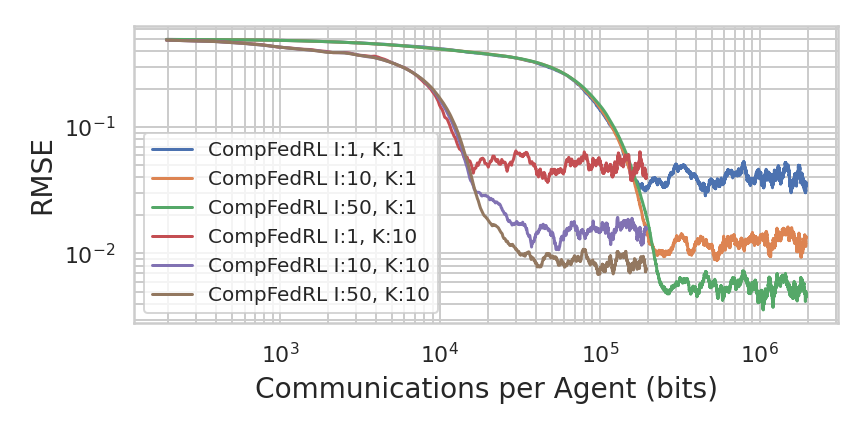}}
  \hfil
\subfloat[\label{speedup_Map1}]{
\includegraphics[width=0.49\textwidth]{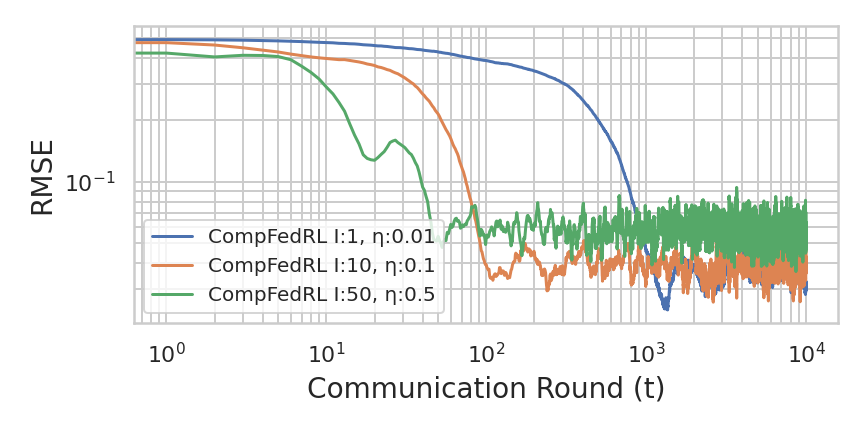}}
\caption{Impact of the number of agents $I$ and the number of local epochs $K$: the RMSE of the $Q$-estimates for \compfedrl ~under \textsf{Top-5} sparsification to solve Map$5\times5$ task.
Here, (a) $T\times K = 10000$, the learning rate $\eta =0.01$, and the federated parameter $\beta=1$.
(b) the number of communication rounds $T= 10000$, the number of local epochs $K=1$, and the federated parameter $\beta=0.8$.}
\label{agent_K_speedup__Map1}
\end{figure*}

\begin{figure*}[!htb] %[H]
\centering
\subfloat[\textsf{Top-5}\label{beta_topk_Map1}]{
  \includegraphics[width=0.49\textwidth]{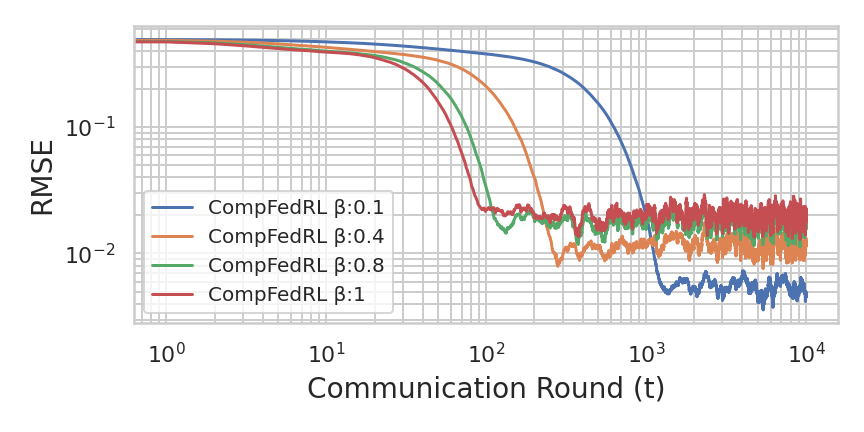}}
  \hfil
\subfloat[\textsf{Sparsified-5}\label{beta_sparsifiedk_Map1}]{
  \includegraphics[width=0.49\textwidth]{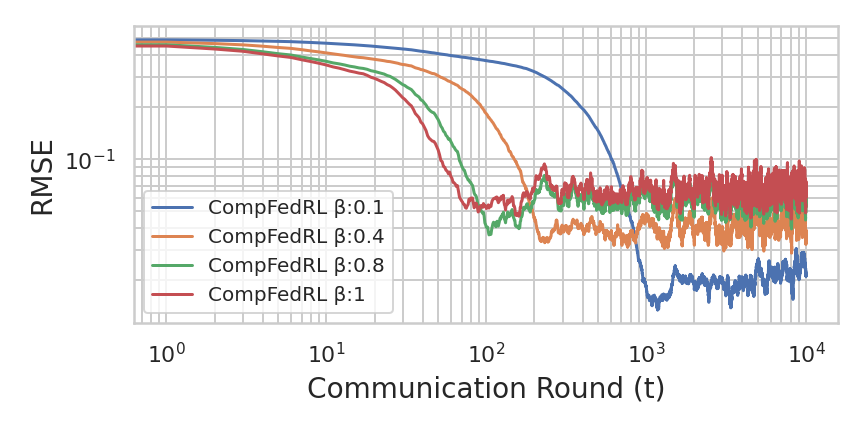}}
\caption{Impact of federated parameter $\beta$: the RMSE of the $Q$-estimates with respect to the communication round for \compfedrl ~under \textsf{Top-5} ~and \textsf{Sparsified-5} ~sparsification to solve Map$5\times5$ task.
Here, the number of agents $I= 50$, the number of communication rounds $T= 10000$, the number of local epochs $K=1$, and the learning rate $\eta =0.1$.}
\label{beta_Map1}
\end{figure*}

\begin{figure*}[!htb] %[H]
\centering
\subfloat[\textsf{Top-5}\label{eta_topk_Map1}]{
  \includegraphics[width=0.49\textwidth]{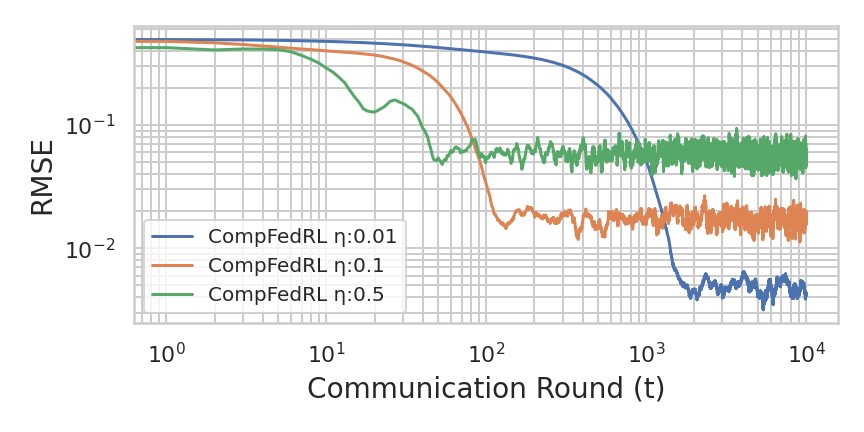}}
  \hfil
\subfloat[\textsf{Sparsified-5}\label{eta_sparsifiedk_Map1}]{
  \includegraphics[width=0.49\textwidth]{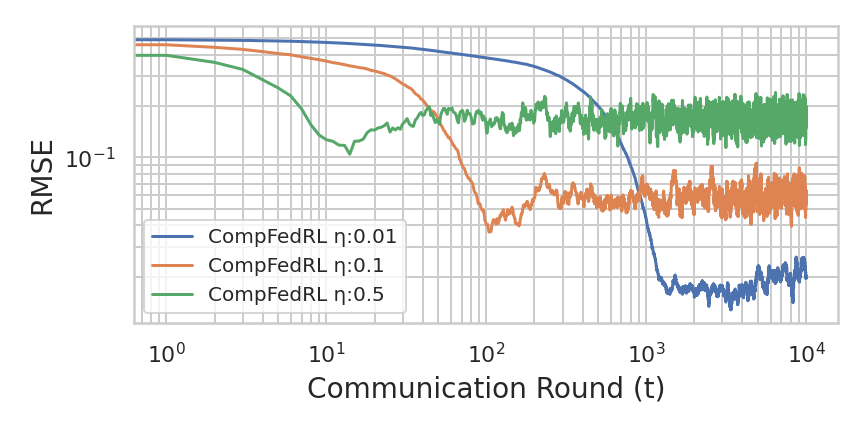}}
\caption{Impact of learning rate $\eta$: the RMSE of the $Q$-estimates with respect to the communication round for \compfedrl ~under \textsf{Top-5} ~and \textsf{Sparsified-5} ~sparsification to solve Map$5\times5$ task. Here, the number of agents $I= 50$, the number of communication rounds $T= 10000$, the number of local epochs $K=1$, and the federated parameter $\beta=0.8$.}
\label{eta_Map1}
\end{figure*}

The effects of compression, the number of agents $I$, and the number of local epochs $K$, as well as the impact of the federated parameter $\beta$ and the learning rate $\eta$ on solving the Map$5\times5$ task, are depicted in Figures \ref{compression_Map1}, \ref{agent_K_speedup__Map1}, \ref{beta_Map1}, and \ref{eta_Map1} respectively. It's noteworthy that due to the smaller state-action spaces in the Map$5\times5$ task compared to the Map$11\times11$ task, more compressed operators, specifically \textsf{Top-5} and \textsf{Sparsified-5}, were utilized instead of \textsf{Top-50} and \textsf{Sparsified-50}.

\subsubsection{Results for Map$5\times5$w task.} \label{apendix:Map2}
\begin{figure*}[!htb] %[H]
\centering
\subfloat[Number of local epochs $K=1$\label{compression_K1_Map2}]{
  \includegraphics[width=0.49\textwidth]{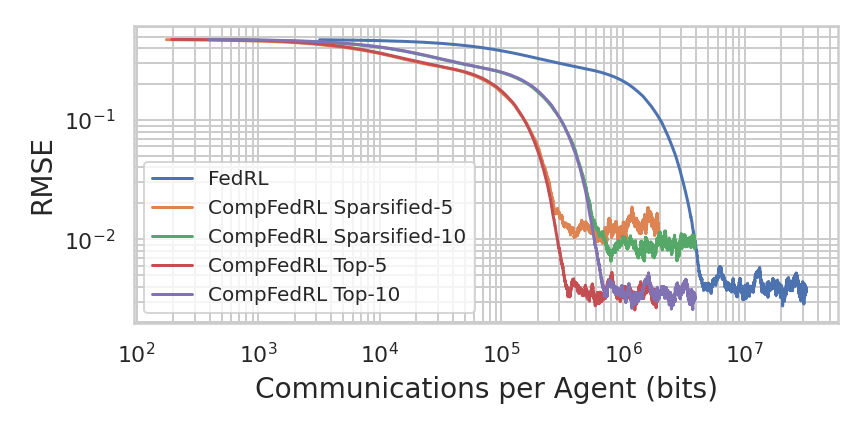}}
  \hfil
\subfloat[Number of local epochs $K=10$\label{compression_K10_Map2}]{
  \includegraphics[width=0.49\textwidth]{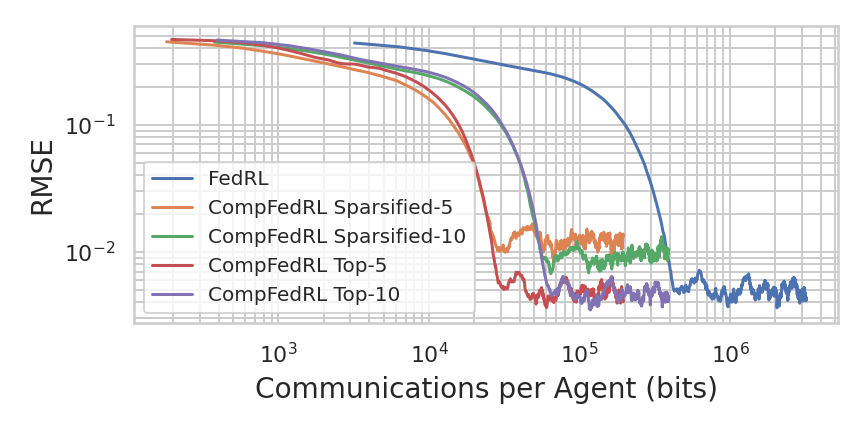}}
\caption{Impact of compression: the RMSE of the $Q$-estimates with respect to the number of bits communicated per agent for both \textsf{FedRL} (without compression) and \compfedrl ~under different levels of sparsification via \sparsifiedk ~and \topk ~to solve Map$5\times5$w task.
Here, the number of agents $I= 50$, the learning rate $\eta =0.01$, and the federated parameter $\beta= 0.8$.}
\label{compression_Map2}
\end{figure*}

\begin{figure*}[!htb] %[H]
\centering
\subfloat[\label{agent_K_Map2}]{
\includegraphics[width=0.49\textwidth]{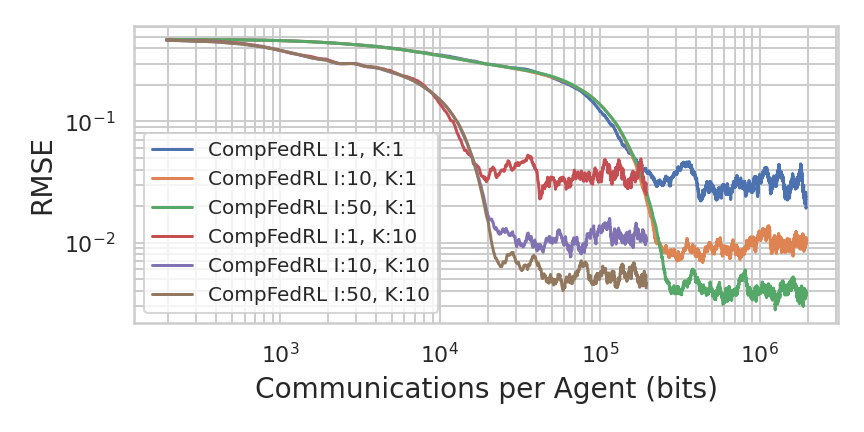}}
  \hfil
\subfloat[\label{speedup_Map2}]{
\includegraphics[width=0.49\textwidth]{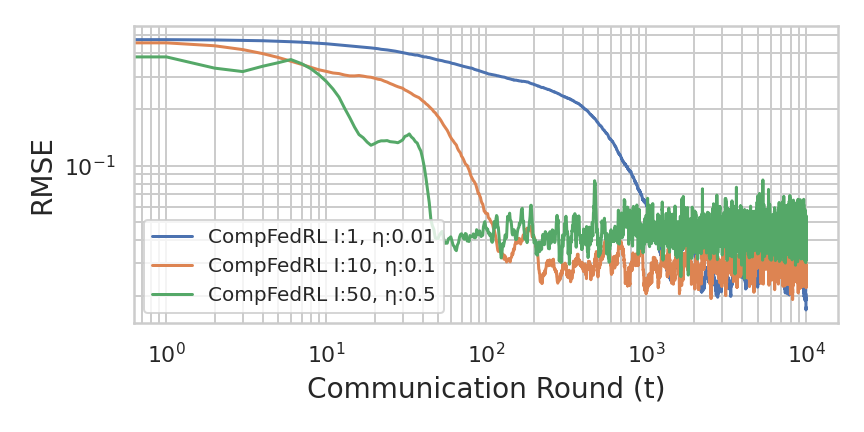}}
\caption{Impact of the number of agents $I$ and the number of local epochs $K$: the RMSE of the $Q$-estimates for \compfedrl ~under \textsf{Top-5} sparsification to solve Map$5\times5$w task.
Here, (a) $T\times K = 10000$, the learning rate $\eta =0.01$, and the federated parameter $\beta=1$.
(b) the number of communication rounds $T= 10000$, the number of local epochs $K=1$, and the federated parameter $\beta=0.8$.}
\label{agent_K_speedup__Map2}
\end{figure*}

\begin{figure*}[!htb] %[H]
\centering
\subfloat[\textsf{Top-5}\label{beta_topk_Map2}]{
  \includegraphics[width=0.49\textwidth]{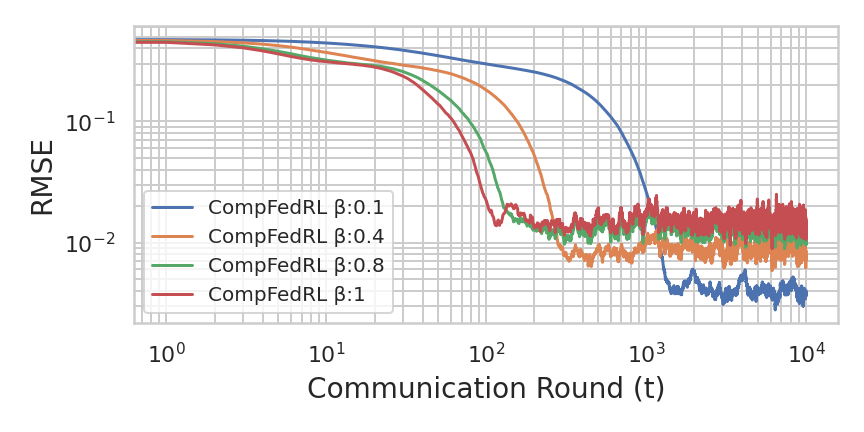}}
  \hfil
\subfloat[\textsf{Sparsified-5}\label{beta_sparsifiedk_Map2}]{
  \includegraphics[width=0.49\textwidth]{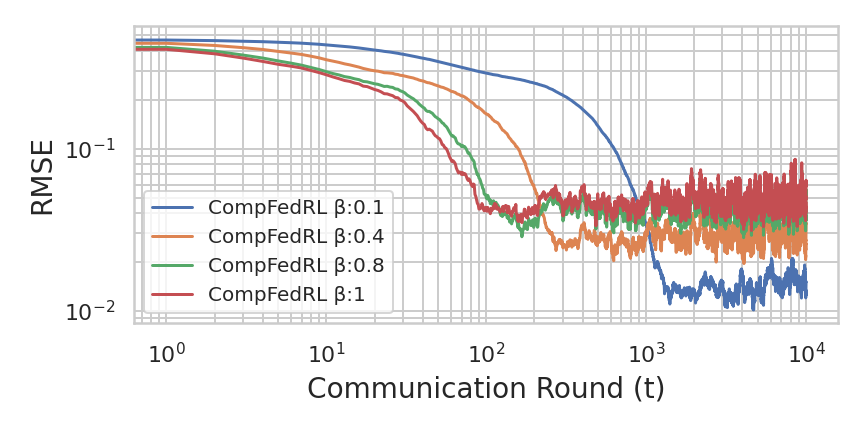}}
\caption{Impact of federated parameter $\beta$: the RMSE of the $Q$-estimates with respect to the communication round for \compfedrl ~under \textsf{Top-5} ~and \textsf{Sparsified-5} ~sparsification to solve Map$5\times5$w task.
Here, the number of agents $I= 50$, the number of communication rounds $T= 10000$, the number of local epochs $K=1$, and the learning rate $\eta =0.1$.}
\label{beta_Map2}
\end{figure*}

\begin{figure*}[!htb] %[H]
\centering
\subfloat[\textsf{Top-5}\label{eta_topk_Map2}]{
  \includegraphics[width=0.49\textwidth]{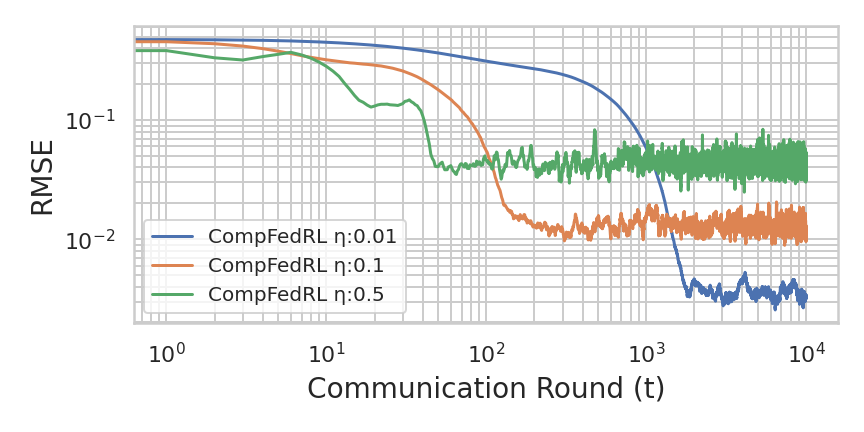}}
  \hfil
\subfloat[\textsf{Sparsified-5}\label{eta_sparsifiedk_Map2}]{
  \includegraphics[width=0.49\textwidth]{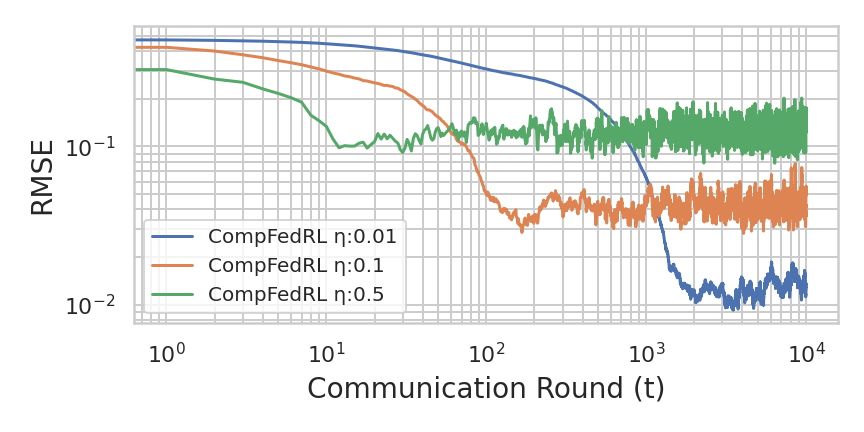}}
\caption{Impact of learning rate $\eta$: the RMSE of the $Q$-estimates with respect to the communication round for \compfedrl ~under \textsf{Top-5} ~and \textsf{Sparsified-5} ~sparsification to solve Map$5\times5$w task. Here, the number of agents $I= 50$, the number of communication rounds $T= 10000$, the number of local epochs $K=1$, and the federated parameter $\beta=0.8$.}
\label{eta_Map2}
\end{figure*}

The impacts of compression, the number of agents \(I\), and the number of local epochs \(K\), alongside the influence of the federated parameter \(\beta\) and the learning rate \(\eta\) on addressing the Map$5\times5$w task, are illustrated in Figures \ref{compression_Map2}, \ref{agent_K_speedup__Map2}, \ref{beta_Map2}, and \ref{eta_Map2}, respectively. It is worth noting that due to the smaller state-action spaces in the Map$5\times5$w task compared to the Map$11\times11$ task, more compressed operators—specifically \textsf{Top-5} and \textsf{Sparsified-5}—were employed instead of \textsf{Top-50} and \textsf{Sparsified-50}, respectively.

\subsubsection{Results for Map$6\times6$w task.} \label{apendix:Map3}
\begin{figure*}[!htb] %[H]
\centering
\subfloat[Number of local epochs $K=1$\label{compression_K1_Map3}]{
  \includegraphics[width=0.49\textwidth]{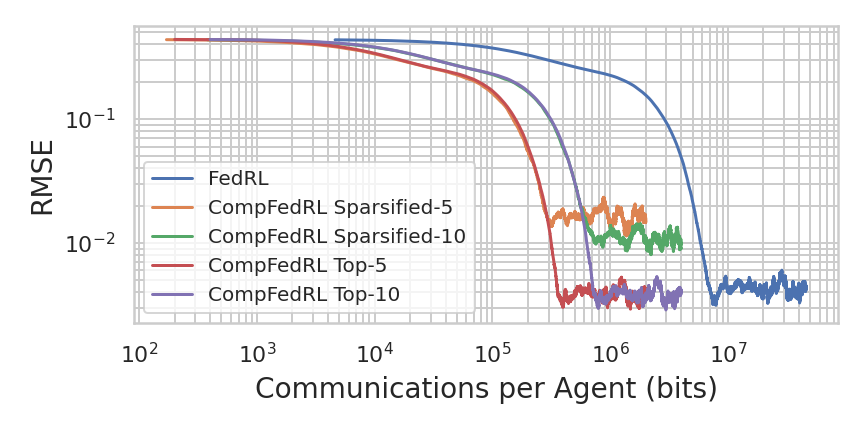}}
  \hfil
\subfloat[Number of local epochs $K=10$\label{compression_K10_Map3}]{
  \includegraphics[width=0.49\textwidth]{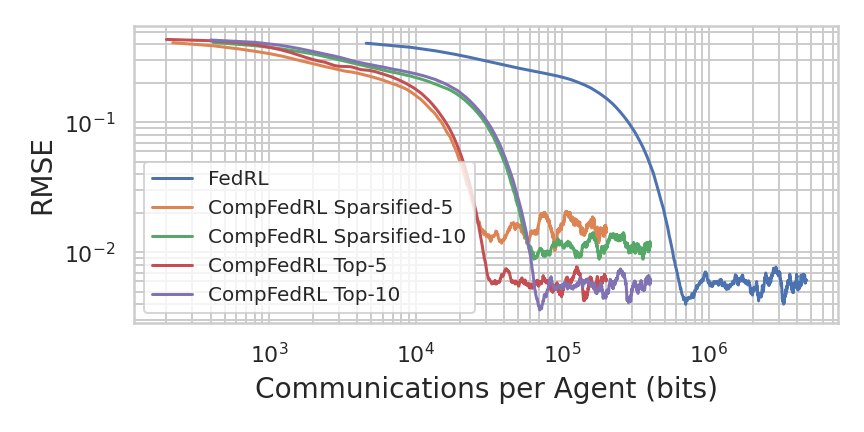}}
\caption{Impact of compression: the RMSE of the $Q$-estimates with respect to the number of bits communicated per agent for both \textsf{FedRL} (without compression) and \compfedrl ~under different levels of sparsification via \sparsifiedk ~and \topk ~to solve Map$6\times6$w task.
Here, the number of agents $I= 50$, the learning rate $\eta =0.01$, and the federated parameter $\beta= 0.8$.}
\label{compression_Map3}
\end{figure*}

\begin{figure*}[!htb] %[H]
\centering
\subfloat[\label{agent_K_Map3}]{
\includegraphics[width=0.49\textwidth]{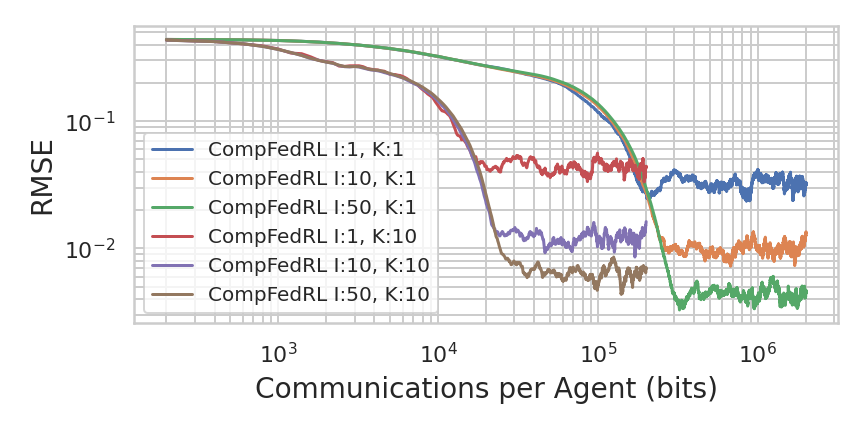}}
  \hfil
\subfloat[\label{speedup_Map3}]{
\includegraphics[width=0.49\textwidth]{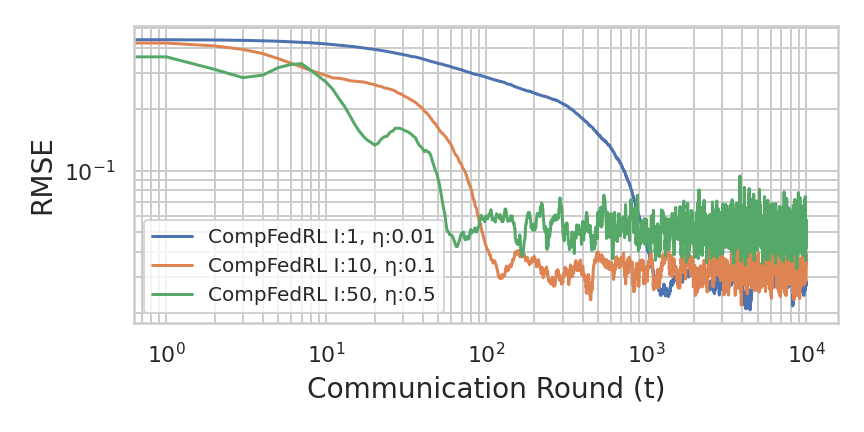}}
\caption{Impact of the number of agents $I$ and the number of local epochs $K$: the RMSE of the $Q$-estimates for \compfedrl ~under \textsf{Top-5} sparsification to solve Map$6\times6$w task.
Here, (a) $T\times K = 10000$, the learning rate $\eta =0.01$, and the federated parameter $\beta=1$.
(b) the number of communication rounds $T= 10000$, the number of local epochs $K=1$, and the federated parameter $\beta=0.8$.}
\label{agent_K_speedup__Map3}
\end{figure*}

\begin{figure*}[!htb] %[H]
\centering
\subfloat[\textsf{Top-5}\label{beta_topk_Map3}]{
  \includegraphics[width=0.49\textwidth]{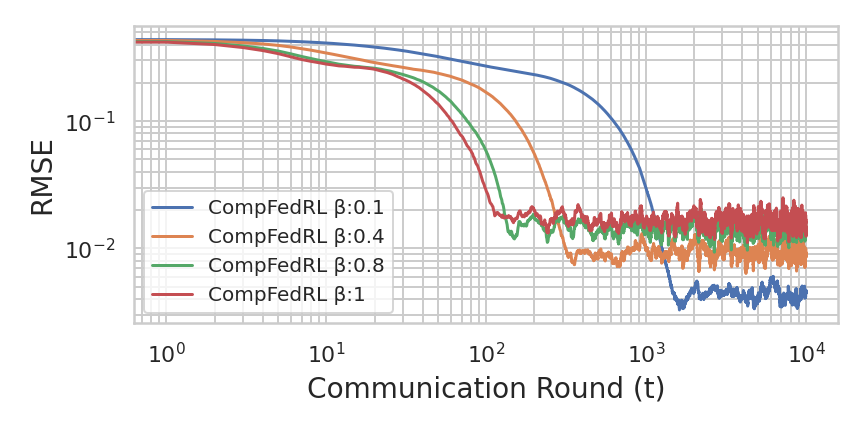}}
  \hfil
\subfloat[\textsf{Sparsified-5}\label{beta_sparsifiedk_Map3}]{
  \includegraphics[width=0.49\textwidth]{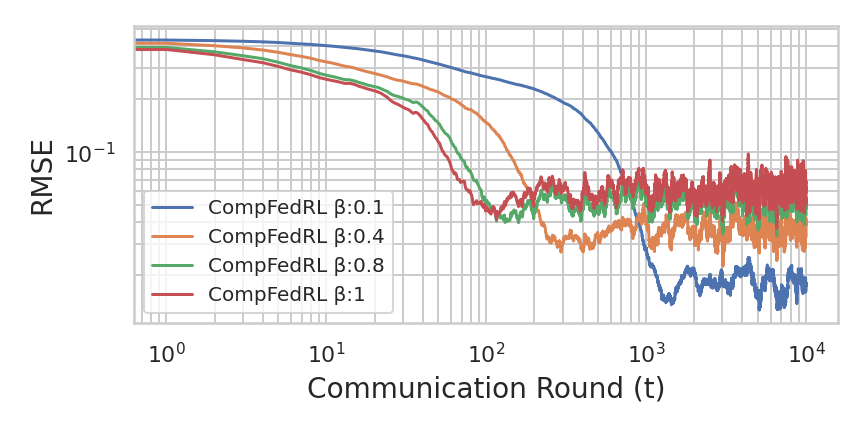}}
\caption{Impact of federated parameter $\beta$: the RMSE of the $Q$-estimates with respect to the communication round for \compfedrl ~under \textsf{Top-5} ~and \textsf{Sparsified-5} ~sparsification to solve Map$6\times6$w task.
Here, the number of agents $I= 50$, the number of communication rounds $T= 10000$, the number of local epochs $K=1$, and the learning rate $\eta =0.1$.}
\label{beta_Map3}
\end{figure*}

\begin{figure*}[!htb] %[H]
\centering
\subfloat[\textsf{Top-5}\label{eta_topk_Map3}]{
  \includegraphics[width=0.49\textwidth]{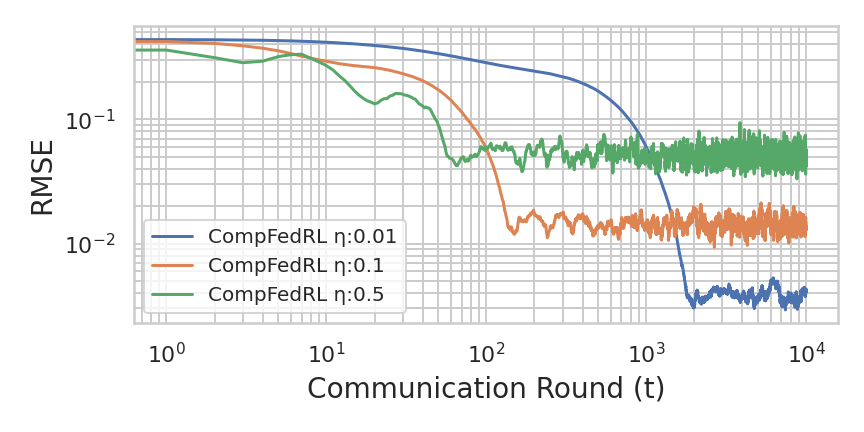}}
  \hfil
\subfloat[\textsf{Sparsified-5}\label{eta_sparsifiedk_Map3}]{
  \includegraphics[width=0.49\textwidth]{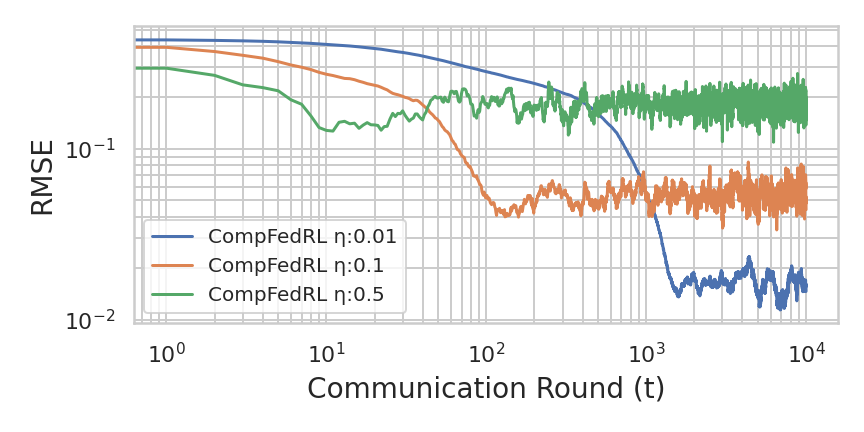}}
\caption{Impact of learning rate $\eta$: the RMSE of the $Q$-estimates with respect to the communication round for \compfedrl ~under \textsf{Top-5} ~and \textsf{Sparsified-5} ~sparsification to solve Map$6\times6$w task. Here, the number of agents $I= 50$, the number of communication rounds $T= 10000$, the number of local epochs $K=1$, and the federated parameter $\beta=0.8$.}
\label{eta_Map3}
\end{figure*}

The impacts of compression, the number of agents \(I\), and the number of local epochs \(K\), along with the influence of the federated parameter \(\beta\) and the learning rate \(\eta\) on addressing the Map$6\times6$w task, are illustrated in Figures \ref{compression_Map3}, \ref{agent_K_speedup__Map3}, \ref{beta_Map3}, and \ref{eta_Map3} respectively. Notably, due to the smaller state-action spaces in the Map$6\times6$w task compared to the Map$11\times11$ task, more compressed operators, specifically \textsf{Top-5} and \textsf{Sparsified-5}, were employed instead of \textsf{Top-50} and \textsf{Sparsified-50}.

\subsubsection{Results for Map$17\times17$w task} \label{apendix:Map5}

\begin{figure*}[!htb] %[H]
\centering
\subfloat[Number of local epochs $K=1$\label{compression_K1_Map5}]{
  \includegraphics[width=0.49\textwidth]{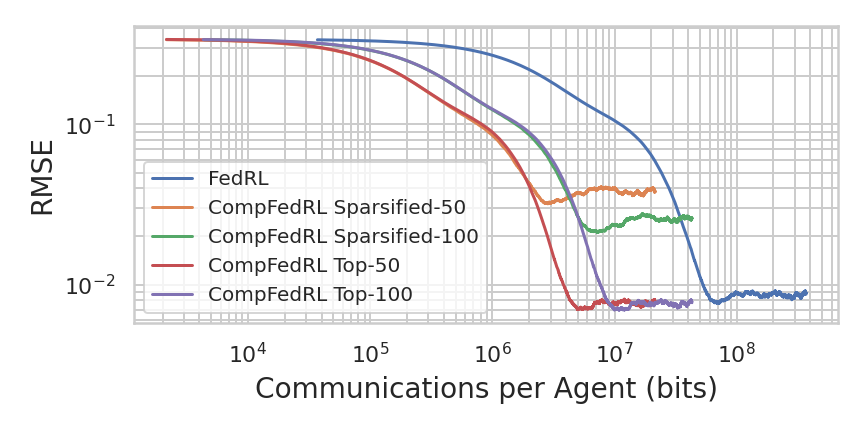}}
  \hfil
\subfloat[Number of local epochs $K=10$\label{compression_K10_Map5}]{
  \includegraphics[width=0.49\textwidth]{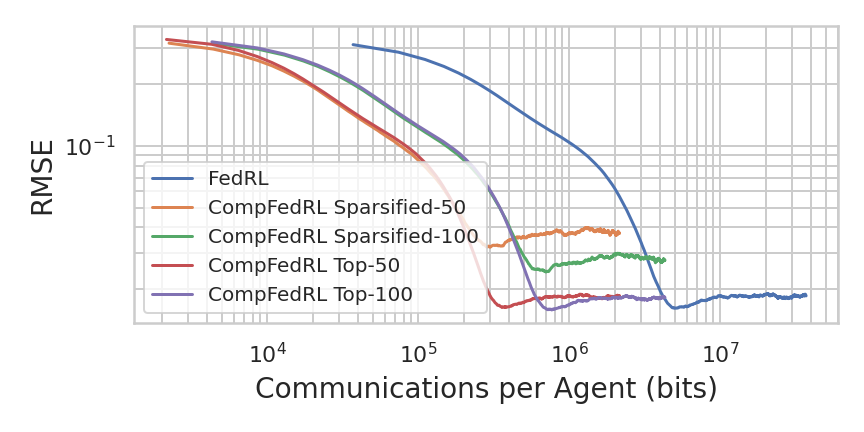}}
\caption{Impact of compression: the RMSE of the $Q$-estimates with respect to the number of bits communicated per agent for both \textsf{FedRL} (without compression) and \compfedrl ~under different levels of sparsification via \sparsifiedk ~and \topk ~to solve Map$17\times17$w task.
Here, the number of agents $I= 50$, the learning rate $\eta =0.01$, and the federated parameter $\beta= 0.8$.}
\label{compression_Map5}
\end{figure*}

\begin{figure*}[!htb] %[H]
\centering
\subfloat[\label{agent_K_Map5}]{
\includegraphics[width=0.49\textwidth]{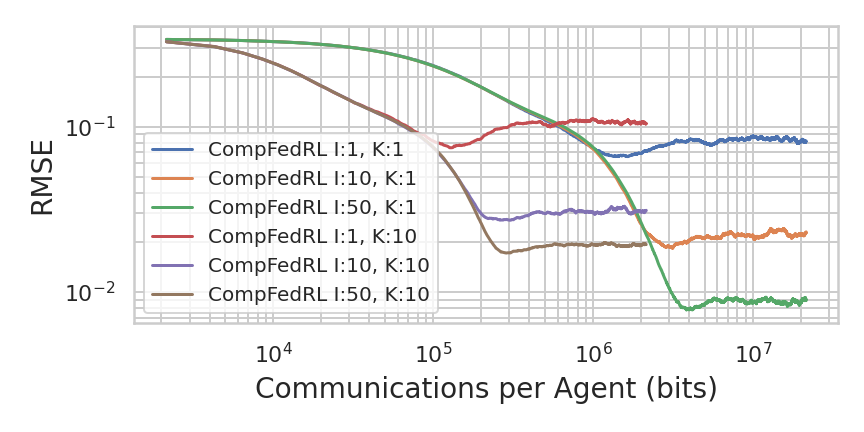}}
  \hfil
\subfloat[\label{speedup_Map5}]{
\includegraphics[width=0.49\textwidth]{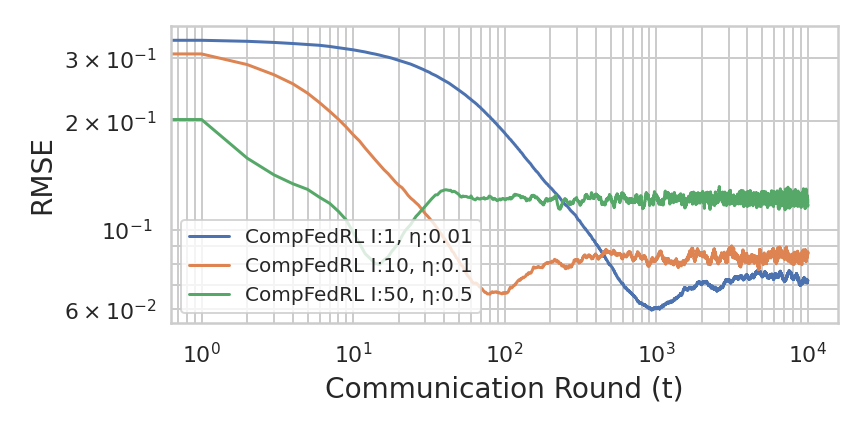}}
\caption{Impact of the number of agents $I$ and the number of local epochs $K$: the RMSE of the $Q$-estimates for \compfedrl ~under \textsf{Top-50} sparsification to solve Map$17\times17$w task.
Here, (a) $T\times K = 10000$, the learning rate $\eta =0.01$, and the federated parameter $\beta=1$.
(b) the number of communication rounds $T= 10000$, the number of local epochs $K=1$, and the federated parameter $\beta=0.8$.}
\label{agent_K_speedup__Map5}
\end{figure*}

\begin{figure*}[!htb] %[H]
\centering
\subfloat[\textsf{Top-50}\label{beta_topk_Map5}]{
  \includegraphics[width=0.49\textwidth]{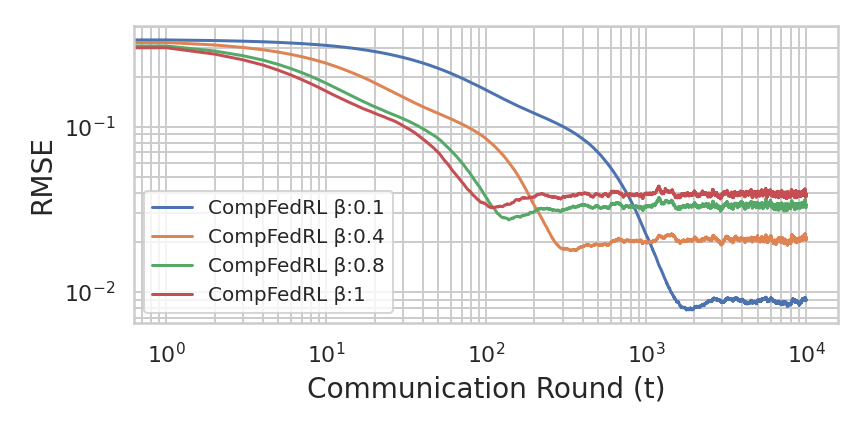}}
  \hfil
\subfloat[\textsf{Sparsified-50}\label{beta_sparsifiedk_Map5}]{
  \includegraphics[width=0.49\textwidth]{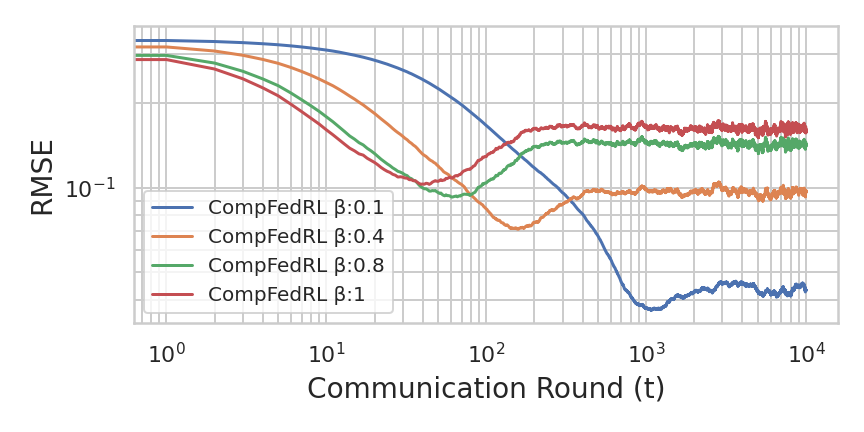}}
\caption{Impact of federated parameter $\beta$: the RMSE of the $Q$-estimates with respect to the communication round for \compfedrl ~under \textsf{Top-50} ~and \textsf{Sparsified-50} ~sparsification to solve Map$17\times17$w task.
Here, the number of agents $I= 50$, the number of communication rounds $T= 10000$, the number of local epochs $K=1$, and the learning rate $\eta =0.1$.}
\label{beta_Map5}
\end{figure*}

\begin{figure*}[!htb] %[H]
\centering
\subfloat[\textsf{Top-50}\label{eta_topk_Map5}]{
  \includegraphics[width=0.49\textwidth]{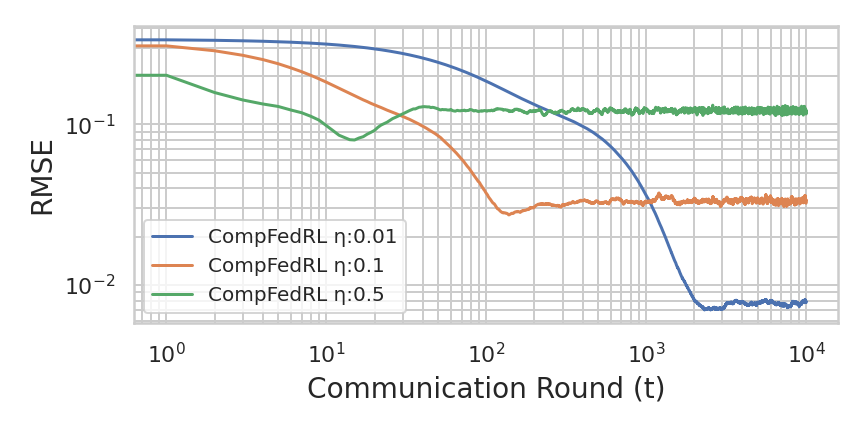}}
  \hfil
\subfloat[\textsf{Sparsified-50}\label{eta_sparsifiedk_Map5}]{
  \includegraphics[width=0.49\textwidth]{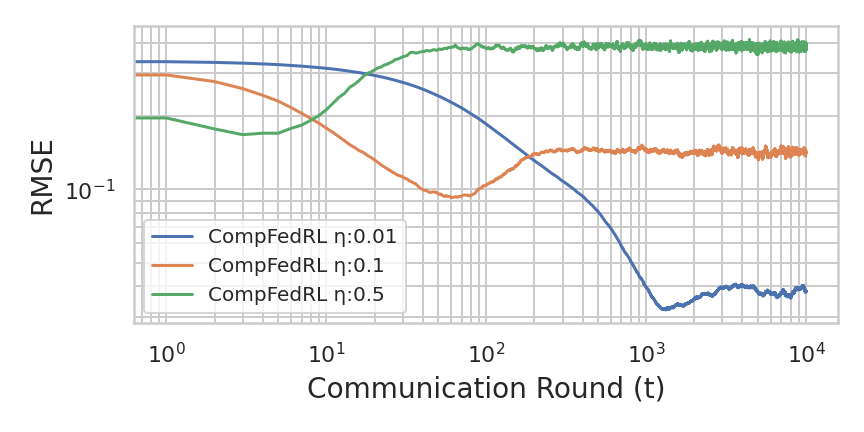}}
\caption{Impact of learning rate $\eta$: the RMSE of the $Q$-estimates with respect to the communication round for \compfedrl ~under \textsf{Top-50} ~and \textsf{Sparsified-50} ~sparsification to solve Map$17\times17$w task. Here, the number of agents $I= 50$, the number of communication rounds $T= 10000$, the number of local epochs $K=1$, and the federated parameter $\beta=0.8$.}
\label{eta_Map5}
\end{figure*}

The impacts of compression, the number of agents \(I\), and the number of local epochs \(K\), along with the influence of the federated parameter \(\beta\) and the learning rate \(\eta\) on solving the Map$17\times17$w task, are illustrated in Figures \ref{compression_Map5}, \ref{agent_K_speedup__Map5}, \ref{beta_Map5}, and \ref{eta_Map5} respectively. 

%\fi

\end{document}